\documentclass[aps,prl,twocolumn,floatfix,superscriptaddress]{revtex4-2}

\usepackage{rotating}
\usepackage{epstopdf}
\usepackage{graphicx}
\usepackage{natbib}
\usepackage{epstopdf}
\usepackage{amsmath}
\usepackage{amssymb}
\usepackage{mathbbol}
\usepackage{xcolor}
\usepackage{lipsum}
\usepackage{mwe}
\usepackage{lipsum} 
\usepackage{siunitx}
\usepackage[normalem]{ulem}
\usepackage[dvipsnames]{xcolor}
\usepackage{cancel}

\usepackage{empheq}
\usepackage{afterpage}
\usepackage[colorlinks,citecolor=blue,urlcolor=red]{hyperref}
\usepackage{cleveref}

\newcommand{\beq}{\begin{equation}}
\newcommand{\eeq}{\end{equation} \smallskip}
\newcommand{\beqy}{\begin{eqnarray}}
\newcommand{\eeqy}{\end{eqnarray} \smallskip}

\newcommand{\bit}{\begin{itemize}}
\newcommand{\eit}{\end{itemize}}
\newcommand{\bmat}{\begin{pmatrix}}
\newcommand{\emat}{\end{pmatrix}}

\makeatletter
\def\maketitle{
\@author@finish
\title@column\titleblock@produce
\suppressfloats[t]}
\makeatother
\newcommand{\beginsupplement}{
    \setcounter{table}{0}
    \renewcommand{\thetable}{S\arabic{table}}
    \setcounter{figure}{0}
    \renewcommand{\thefigure}{S\arabic{figure}}
    \setcounter{equation}{0}
    \setcounter{section}{0}
    \renewcommand{\theequation}{S\arabic{equation}}
}

\begin{document}

\title{Dissipation-enhanced vortex clustering in a compressible quantum fluid}

\author{P. Comaron}
\email[Corresponding author: ]{paolo.comaron@cnr.it} 
\address{CNR NANOTEC, Institute of Nanotechnology, Via Monteroni, 73100 Lecce, Italy}

\author{M. Matuszewski}
\address{Institute of Physics, Polish Academy of Sciences, Al. Lotników 32/46, 02-668 Warsaw, Poland}
\address{Center for Quantum-Enabled Computing, Center for Theoretical Physics, Polish Academy of Sciences, Al. Lotników 32/46, 02-668 Warsaw, Poland}

\author{D. Sanvitto}
\address{CNR NANOTEC, Institute of Nanotechnology, Via Monteroni, 73100 Lecce, Italy}

\author{D. Ballarini}
\address{CNR NANOTEC, Institute of Nanotechnology, Via Monteroni, 73100 Lecce, Italy}

\author{A. S. Lanotte}
\address{CNR NANOTEC, Institute of Nanotechnology, Via Monteroni, 73100 Lecce, Italy}
\address{INFN, Sez. Lecce, 73100 Lecce, Italy}

\begin{abstract}
 Vortex clustering is commonly associated with conservative two-dimensional quantum-fluid dynamics. Therefore, particle loss is mainly expected to limit clustering by shortening the time available for vortex correlations to develop. Here we show instead that particle loss can enhance transient vortex clustering beyond the conservative evolution. We numerically study freely decaying, confined two-dimensional condensates initialized with random distributions of vortices and antivortices, and find a pronounced nonmonotonic dependence of the maximum clustering on particle lifetime. The enhancement is strongest at intermediate particle-loss rates, where loss-induced background rarefaction and incompressible kinetic energy relaxation occur on comparable timescales. Our results identify a finite dynamical window, selected by linear particle loss, in which a confined compressible quantum fluid develops stronger same-sign vortex correlations than in the conservative limit.
\end{abstract}

\maketitle

Two-dimensional (2D) turbulence is distinguished by the emergence of large-scale coherent vortex structures, in contrast to the cascade phenomenology of three-dimensional turbulence~\cite{Frisch1995, kraichnan1967inertial, onsager1949statistical, BEreview}. In 2D quantum fluids, this tendency towards self-organization is closely connected to Onsager vortex clustering, where an initially disordered distribution of vortices and antivortices evolves toward long-lived same-sign aggregates and negative-temperature vortex states~\cite{onsager1949statistical, Groszek2018, simula2014prl}. This physics has attracted considerable interest in atomic BECs~\cite{johnstone2019evolution,gauthier2019giant} and, more recently, in light-matter systems such as exciton-polariton condensates and paraxial quantum fluids of light~\cite{Berloff2010Turbulence, Koniakhin2020CSF, panico2023, Depaepe2026Counterflow, BakerRasooli2023PRA}, where quantized vortices can be directly generated and tracked during their nonlinear evolution.

In weakly dissipative two-dimensional quantum fluids, vortex clustering is commonly understood in terms of evaporative heating. The preferential annihilation or removal of tightly bound vortex--antivortex pairs increases the incompressible energy available per surviving vortex, driving the vortex gas toward negative-temperature states and large-scale same-sign clustering~\cite{Groszek2018,simula2014prl,Valani_2018,Billam2014}. This picture is naturally captured by point-vortex descriptions, which reproduce the long-range configurational dynamics of well-separated vortices and the infrared region of the incompressible kinetic energy spectrum~\cite{Skipp2023,tattersall2025out,Billam2015}. Previous studies based on damped Gross–Pitaevskii models considered energy relaxation associated with a thermal reservoir, which introduces a direct dissipative drift in the vortex dynamics~\cite{Billam2015}. By contrast, the uniform linear particle loss considered here primarily rarefies the compressible background and does not correspond to direct vortex friction. Within this framework, particle loss would be expected mainly to interrupt the evaporative-heating process before correlations can fully develop, suggesting that vortex ordering should improve monotonically toward the conservative limit. 

Uniform linear particle loss, however, does more than remove particles or shorten the available evolution time. In a compressible quantum fluid, it leads to a progressive depletion of the condensate density, thereby changing the sound velocity, healing length, and vortex-depletion dynamics. Vortex motion is also accompanied by sound emission, density redistribution, and vortex creation and annihilation processes~\cite{barenghi_skrbek_sreenivasan_2023,Berloff2004Rarefaction,kobayashi2007quantum,overeng2025topological,navon2016emergence,Panico2025}, which couple the evolution of the vortex gas to that of the background fluid~\cite{Bagnato2020,Groszek2020,Kanai2021,muller2024,tattersall2025out}. In a freely decaying condensate, vortex reorganization therefore occurs concurrently with evaporative heating, vortex depletion, and loss-induced background rarefaction. Particle loss may consequently do more than truncate the conservative evolution: by changing the relative timescales of these coupled processes, it may alter the maximum ordering reached during the finite dynamics. Although the dependence of clustering on particle lifetime has previously been investigated in decaying polariton quantum fluids~\cite{comaron2025clustering, Ferrini2025}, an enhancement beyond the conservative limit was not identified.

Here we show that dissipation can indeed enhance vortex clustering beyond the conservative limit. We study a trapped, freely decaying compressible quantum fluid using parameters representative of exciton-polariton condensates~\cite{panico2023,Estrecho2019, conformal2023}, while varying the dissipation rates across a broader weak-dissipation regime relevant to compressible quantum fluids more generally. By scanning the particle lifetime $\tau_\gamma$ over more than two orders of magnitude, we uncover a pronounced nonmonotonic dependence of the maximum clustering. At short lifetimes, ordering is particle-loss limited because the condensate decays before vortex correlations can fully develop. Toward the conservative limit, clustering instead becomes limited by depletion of the vortex population. The strongest ordering occurs at intermediate dissipation, where background rarefaction and incompressible kinetic energy relaxation remain comparable over a finite dynamical window. Moderate particle loss can therefore select a more favorable route to Onsager ordering than the conservative evolution in a finite compressible quantum fluid.

Turbulent dissipative dynamics is modeled by numerically solving the two-dimensional Gross-Pitaevskii equation in the presence of linear particle losses~\cite{Proukakis_Snoke_Littlewood_2017,Barenghi_book},
\begin{equation}
i\hbar\partial_t\psi=
\left[
-\frac{\hbar^2\nabla^2}{2m}+g|\psi|^2+V(\mathbf r)-i\hbar\frac{\gamma}{2}
\right]\psi,
\label{eq:dgpe}
\end{equation}
where $m$ is the effective mass, $g$ the interaction strength, and $\gamma=\tau_\gamma^{-1}$ the particle-loss rate; the conservative limit corresponds to $\gamma=0$. We consider a confined geometry with a radially symmetric potential $V({\mathbf r})$ of diameter $D$, implemented as a hard-bounded box. We initialize the system with a homogeneous condensate in which a spatially random distribution of vortices and antivortices is imprinted, following the standard setup used to study Onsager clustering in two-dimensional quantum fluids~\cite{johnstone2019evolution,gauthier2019giant,panico2023,comaron2025clustering}. By varying, at $t=0$, the condensate density $n_0 \equiv |\psi(t=0)|^2$ and the system size $D$, while keeping the number of vortices $N_{\mathrm v}$ fixed, we independently control the healing length $\xi_0=\hbar/\sqrt{2mgn_0}$ and the intervortex spacing $\ell_{\mathrm v}^0=D/\sqrt{N_{\mathrm v}^0}$, which parameterize the vortex-gas configuration in the initial condition. For each choice $({\xi_0,\ell_{\mathrm v}^0})$ in the parameter space, we average results over a sample of $\mathcal{N} = 5\times 10^2$ numerical simulations starting from random and independent realizations of the initial vortex configurations ~\cite{comaron2025clustering}. 
This ensures high statistical convergence and reduces error bars. Full definitions and numerical details are reported in Ref.~\cite{SI}.

\begin{figure}
    \centering
    \includegraphics[width=\columnwidth]{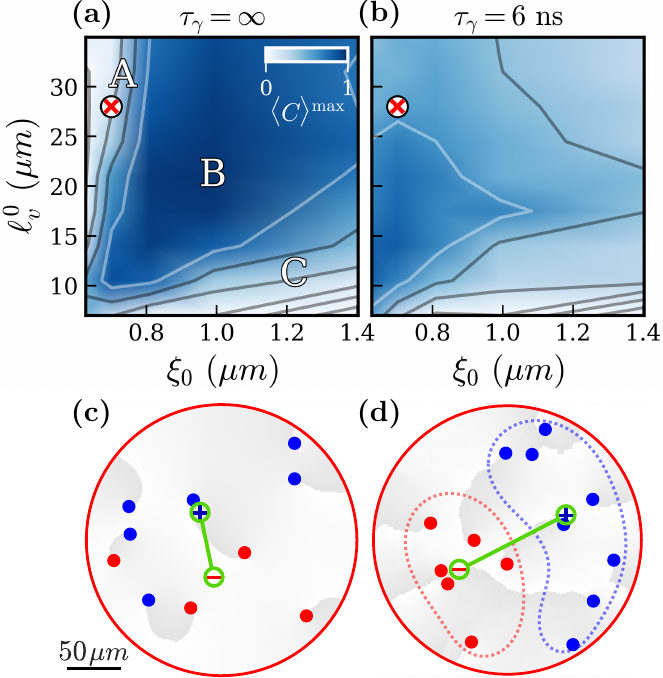}
	\caption{
    Phase portrait of vortex clustering showing the maximum of the correlation $C_\mathrm{max}\equiv \max (\langle C \rangle)$ as a function of the initial values of the intervortex distance $\ell_\mathrm{v}^0$ and healing length $\xi_0$. The average $\langle \bullet \rangle$ is taken over the statistical sample.
    Panel \textbf{(a)} refers to the conservative case ($\tau_{\gamma} \to \infty$), while panel \textbf{(b)} refers to a dissipative case with $\tau_{\gamma}= 6 \mathrm{ns}$. The continuous lines fix the level sets: the white line delimiting region ``B'' is the locus of $C_\mathrm{max}=0.85$, while the others are level sets separated by $0.15$ difference. The red-crossed filled circles correspond to the case with $(\xi_0,\ell^0_\mathrm{v}) = (0.7,28) \mathrm{\mu m}$, and an initial sound velocity of $c_s = \sqrt{gn/m} = 3.02 \mathrm{\mu m/ps}$ (see text). 
    Panels \textbf{(c)} and \textbf{(d)} show real-space snapshots at t=20ns for this parameter set in the conservative and dissipative cases, respectively. Vortices and antivortices (blue and red dots) are superimposed on the condensate phase map, displayed in grayscale. The green line corresponds to the dipole moment $d = N_\mathrm{v}^{-1}|\Sigma_{i=1}^{N_\mathrm{v}} \mathrm{sgn}(\Gamma_i) r_i|$. 
    Data are obtained by setting $m = 0.22 \ \si{ps^2 meV \mu m^{-2}}$, and $g = 5  \si{\mu eV \mu m^2}$.
    }
	\label{fig:phase_diagrams}
\end{figure}

We start by considering the clustering efficiency as quantified by the maximum value of the mean vortex correlation, $C_{\max} \equiv \max({\langle C \rangle})$. The correlation function is defined as $C = N_\mathrm{v}^{-1}
\sum_{i=1}^{N_\mathrm{v}}c_i$, where $c_i$ is the product of the (adimensional) circulations between the $i$-th vortex and its nearest neighbor \cite{gauthier2019giant}. Starting from $C \approx 0$ for a random vortex distribution, values approaching $C=1$ indicate the development of nearest-neighbor correlations between same-sign vortices. 
In the simplest situation of the conservative limit, Figure~\ref{fig:phase_diagrams}(a) shows that strong clustering is achieved over a broad region of the parameter space, region ``B'', consistent with the tendency of a two-dimensional vortex gas to evolve toward Onsager-like same-sign aggregates \cite{Billam2014}.
Nevertheless, two regions of reduced clustering can be identified. For large initial intervortex spacing $\ell_{\mathrm v}^0$ and small $\xi_0$, region ``A'', the vortex gas is too dilute relative to the system size to sustain efficient large-scale organization, while for very small $\ell_{\mathrm v}^0$ and large $\xi_0$, region ``C'', rapid early vortex-antivortex annihilation suppresses the buildup of long-lived ordered structures. 

In the dissipative case, Fig.~\ref{fig:phase_diagrams}(b), similar trends remain present, as revealed by the clustering isocurves, but a qualitatively different behavior emerges at small healing lengths. Here, dissipation enhances clustering beyond the conservative case, and the enhancement is most pronounced for intermediate-to-large initial intervortex spacing, as also illustrated by the representative phase snapshots in Figs.~\ref{fig:phase_diagrams}(c,d). These correspond to the parameter set marked in panels (a,b), with vortices (blue), antivortices (red), and the dipole moment (green) $d = N_\mathrm{v}^{-1}|\Sigma_{i=1}^{N_\mathrm{v}} \mathrm{sgn}(\Gamma_i) r_i|$ superimposed. Here, $\Gamma_i$ and $r_i$ correspond to the circulation and position of the $i$-th vortex, respectively.

\begin{figure}[]
    \centering
    \includegraphics[width=\columnwidth]{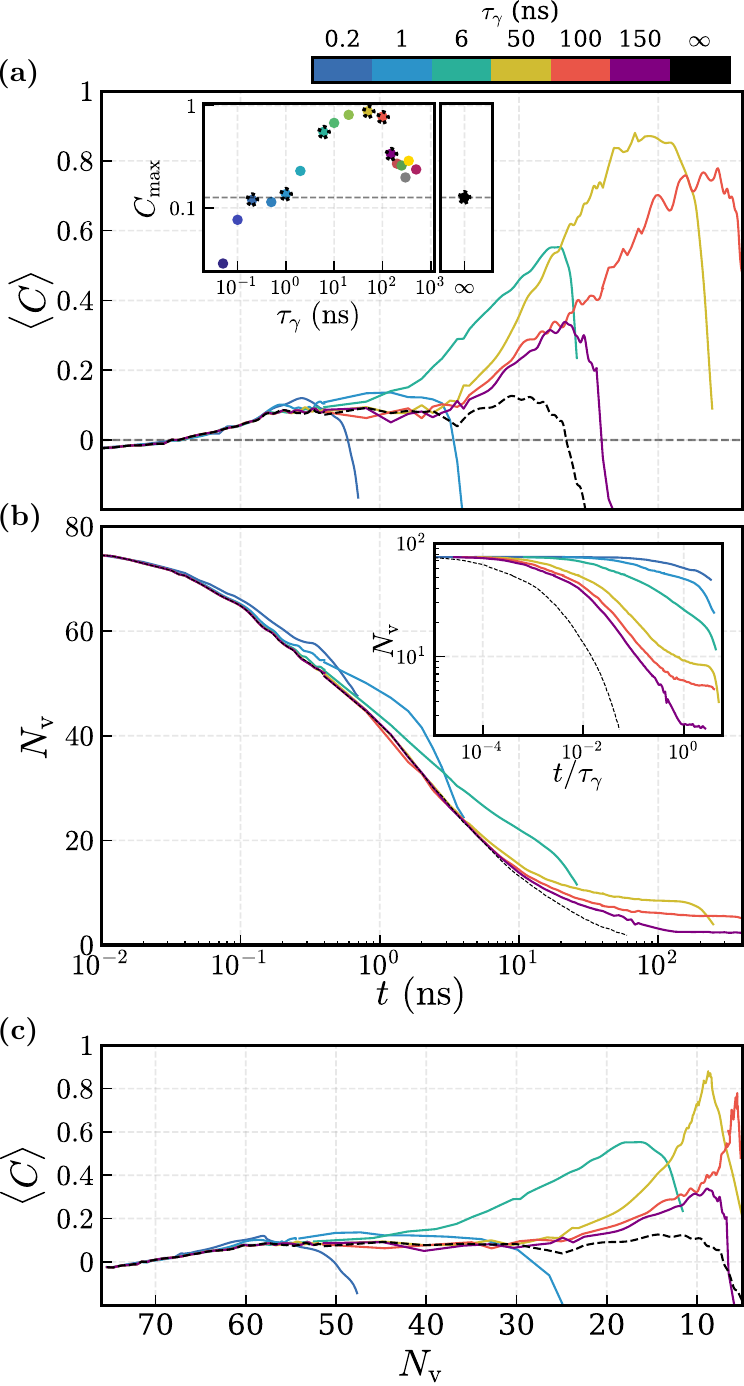}
    \caption{\textbf{(a)} 
    The temporal evolution of the correlation $\langle C \rangle$, for condensates with particle lifetime in the range $\tau_{\gamma} \in [0.2 -150]$ns. In the inset, the maximal clustering value $C_\mathrm{max}$ is reported as a function of the particle lifetime. Statistical error bars are smaller than the symbol size. 
    \textbf{(b)} The temporal evolution of the vortex number $N_\mathrm{v}(t)$ is plotted as a function of time (main panel) and as a function of the rescaled time (inset).
    \textbf{(c)} Parametric representation of $\langle C(t)\rangle$ as a function of the remaining vortex population $N_{\mathrm v}(t)$, with time progressing from left to right. The distinct trajectories show that the relation between clustering and vortex depletion depends on the particle lifetime.
    The conservative evolution is reported as a black dashed line in all panels and insets. In panel \textbf{(a)} and in the inset of panel \textbf{(b)}, the time axis is not made dimensionless for the conservative case since $\tau_\gamma=\infty$.}
	\label{fig:evolution_C_Nv}
\end{figure}

In the following, we focus on a specific system configuration within the dissipation-enhanced region, identified by 
the choice $(\xi_0,\ell_{\mathrm v}^0)=(0.7,28)\,\mathrm{\mu m}$ (see 
the red crossed circle in Fig.~\ref{fig:phase_diagrams}), and vary the particle lifetime over more than two orders of magnitude. 
To determine the nature of the change in the dissipative regime, we examine the full temporal evolution of the vortex correlation function, depicted in  Fig.~\ref{fig:evolution_C_Nv}(a). Numerical results show a pronounced non-monotonic dependence of the clustering kinetics on the particle lifetime $\tau_\gamma$. For very short lifetimes, the condensate decays before correlations can fully build up and clustering is reduced with respect to the conservative case. 
As the lifetime increases, a distinct optimal regime appears, in which $\langle C\rangle$ not only has the time to develop, but it reaches values exceeding the conservative limit (dashed black line). For even larger lifetimes, the maximal correlation function $C_\mathrm{max}$ decreases again, approaching the conservative behavior. The inset of Fig.~\ref{fig:evolution_C_Nv}(a) makes this nonmonotonic trend explicit, showing a clear maximum of $C_{\max}$ at intermediate dissipation values 
\footnote{Note that $C_{\max}$ is always attained with at least approximately ten vortices still remaining, and that the enhancement with respect to the conservative case largely exceeds the statistical uncertainty of the ensemble average.}.

The competing roles of particle loss and vortex depletion are examined in Fig.~\ref{fig:evolution_C_Nv}(b), which shows that particle loss slows the decay of the vortex population. To determine whether the enhanced clustering merely results from a rescaling of the same evolution, Fig.~\ref{fig:evolution_C_Nv}(c) plots $\langle C(t)\rangle$ parametrically against $N_{\mathrm v}(t)$. The curves do not collapse: for the same remaining vortex population, intermediate-lifetime condensates develop stronger same-sign correlations than both the short-lived and conservative cases.

The temporal evolution of the vortex population and clustering correlations must be complemented by energetic considerations. We therefore examine the decay of the total kinetic energy, and of its incompressible component, extracted through the Helmholtz decomposition of the hydrodynamic density-weighted velocity~\cite{SI}. The resulting decomposition shows that the kinetic-energy is dominated by the compressible component throughout the dynamics (see~\cite{SI} for details).

\begin{figure}[t]
    \centering
    \includegraphics[width=\columnwidth]{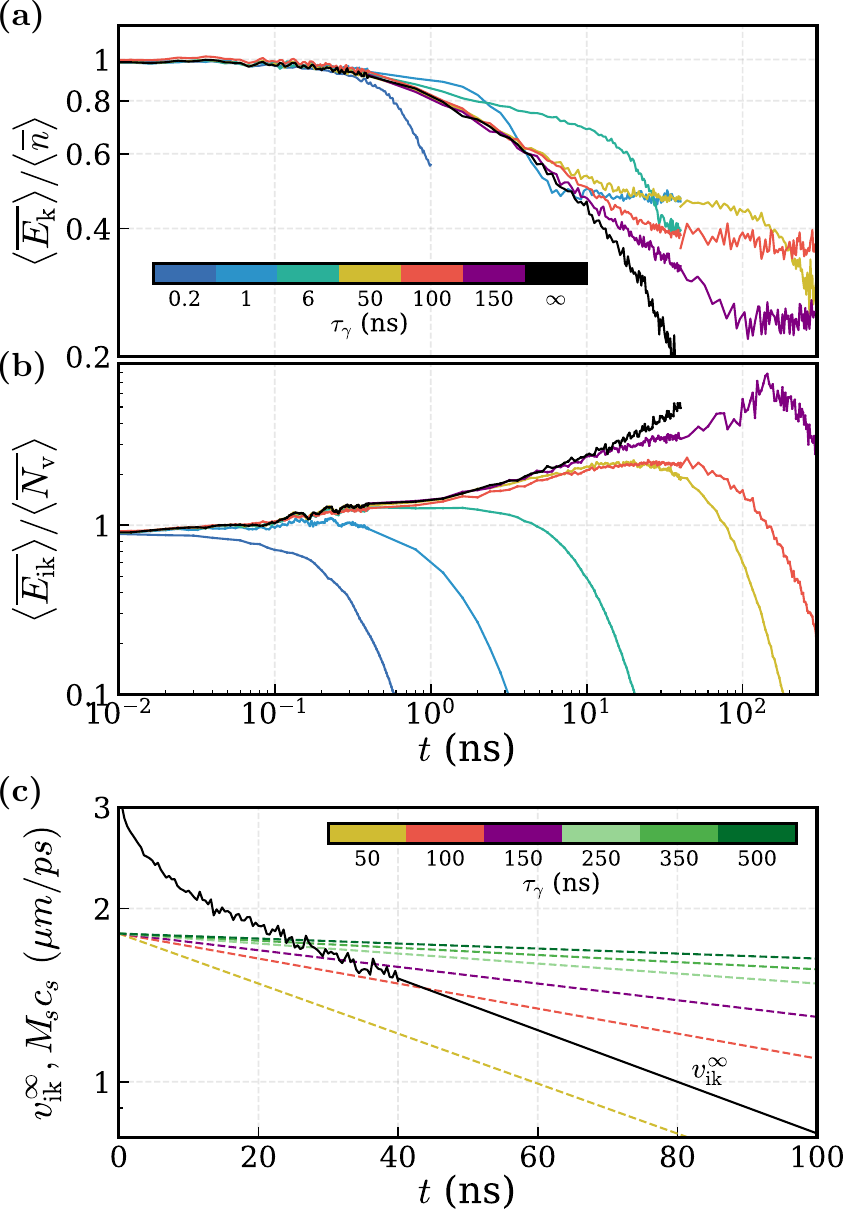}
    \caption{\textbf{(a)} Time evolution of the normalised total kinetic energy per particle $\langle \overline{E_{\mathrm{k}}} \rangle/\langle \overline{n} \rangle$, where $\langle \overline{E_{\mathrm{k}}} \rangle \equiv  {\langle E_{\mathrm{k}}(t)\rangle}/{\langle E_{\mathrm{k}}(t=0)\rangle}$ and $\langle \overline{n} \rangle \equiv {\langle n(t) \rangle}/{n_0}$ (in the conservative case, $\langle \overline{n} \rangle =1$ by definition). \textbf{(b)} Time evolution of the normalised incompressible kinetic energy per vortex, $\langle \overline{E_{ik}} \rangle/ \langle \overline{N_\mathrm{v}} \rangle$. \textbf{(c)} The comparison between the characteristic velocity scale of the vortex sector and the sound-velocity scale associated with particle-loss-induced rarefaction, from which the crossover time $t_s$ reported in Fig.~\ref{fig:Cmax_timescales} is extracted. The value of $M_s=0.6$ is defined to match the crossing time $t_s$ with the depletion time in the conservative case}    \label{fig:evolution_energy_particles_velocity}
\end{figure}

The physical mechanism becomes clearer from Fig.~\ref{fig:evolution_energy_particles_velocity}, which separates the roles of the total kinetic energy, the vortex sector, and the crossover toward the conservative limit. Figure~\ref{fig:evolution_energy_particles_velocity}(a) shows the decay of the total kinetic energy per particle, i.e., comprising both the compressible and incompressible sectors, which follows the same non-monotonic trend found in the clustering dynamics. 
For short particle lifetimes, i.e., $\tau_{\gamma}\le 1\,$ns, the normalized kinetic energy decays faster than in the conservative case, showing that the condensate dynamics are directly truncated by particle losses. However, for intermediate lifetimes, the decay of the kinetic energy per particle becomes slower than in the conservative case, before eventually approaching the conservative behavior again as $\tau_\gamma\to\infty$. Therefore, Fig.~\ref{fig:evolution_energy_particles_velocity}(a) already reveals that dissipation does not simply rescale the conservative dynamics, but promotes a competition between different timescales.

The complementary information from the incompressible sector is shown in Fig.~\ref{fig:evolution_energy_particles_velocity}(b), where we plot the incompressible kinetic energy per vortex, $\langle E_{\mathrm{ik}}\rangle/\langle N_v\rangle$. This quantity increases systematically as the dynamics approach the conservative limit, consistently with the familiar evaporative-heating picture. Importantly, however, this increase does not translate directly into stronger ordering. In contrast, maximal clustering is obtained in the regime where $\langle E_{\mathrm{ik}}\rangle/\langle N_v\rangle$ remains approximately stationary over a finite time window. In other words, the bare increase in incompressible energy available per surviving vortex does not by itself determine the maximum same-sign vortex correlation. Evaporative heating is the natural ordering mechanism in reduced vortex-gas descriptions, where the total incompressible energy is approximately conserved. In the present compressible and dissipative fluid, by contrast, vortex depletion and background rarefaction evolve together, so ordering is controlled by their relative timescales rather than by vortex removal alone.

This point is made explicit in Fig.~\ref{fig:evolution_energy_particles_velocity}(c), which addresses the nontrivial side of the phase diagram, namely, why clustering decreases again as the particle lifetime is increased beyond the point of maximal enhancement. We therefore focus on the long-lifetime side of the dynamics, i.e., from the clustering maximum toward the conservative limit. As a reference for the intrinsic relaxation of the vortex sector, we use the root-mean-square incompressible kinetic velocity in the conservative case,
\begin{equation}
v_{\mathrm{ik}}^\infty(t)=\sqrt{2\langle E_{\mathrm{ik}}^\infty\rangle/m},    
\end{equation}
and compare it with the dissipative sound-velocity curves, proportional to $c_s(t)$. The strongest enhancement of vortex correlation is observed for particle lifetimes such that the decay of $c_s(t)$ becomes approximately parallel to that of $v_{\mathrm{ik}}^\infty(t)$, indicating that background rarefaction and  incompressible kinetic energy relaxation occur on comparable timescales.

This link is made quantitative by noting that the crossing points in Fig.~\ref{fig:evolution_energy_particles_velocity}(c) follow the same trend as the vortex-cloud depletion time, $t_\mathrm{vcd}$, i.e., the time at which the vortex population is completely depleted. In the numerical simulations, this corresponds to the first instant at which no vortices remain in the system.

\begin{figure}
    \centering
    \includegraphics[width=\columnwidth]{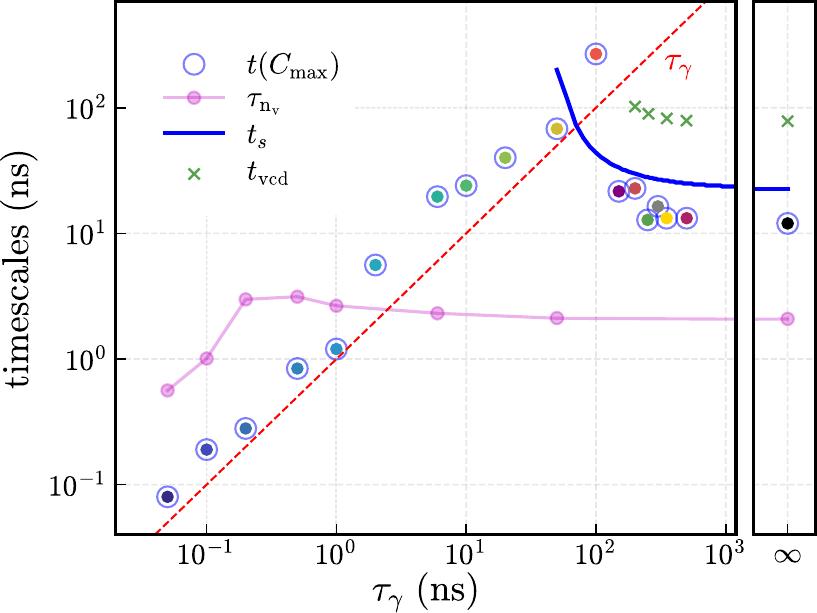}
    \caption{The time of maximum clustering \(t(C_{\mathrm{max}})\) (filled circles) is compared to the particle lifetime \(\tau_\gamma\) (red dashed line), to the characteristic vortex-depletion timescale \(\tau_{n_\mathrm{v}}\) (magenta line with dots), and to the crossover times \(t_s\) (blue line) extracted from Fig.~3(c), where the characteristic velocity scale of the vortex sector becomes comparable to the sound-velocity scale associated with particle-loss-induced rarefaction. Operationally, \(\tau_{n_\mathrm{v}}\) marks the onset of the regime in which vortex correlations begin to compete with pure particle-loss truncation, while \(t_s\) marks the opposite boundary beyond which the residual vortex population no longer sustains efficient growth of clustering. For cases with lifetimes longer than 100 ns, green crosses denote the vortex-cloud dilution time ($t_\mathrm{vcd}$) at which the vortex cloud is fully diluted. Altogether, these results identify a finite dynamical window for optimal clustering between a particle-loss-limited regime and a vortex-loss-limited regime.}
	\label{fig:Cmax_timescales}
\end{figure}

We summarize these observations in Fig.~\ref{fig:Cmax_timescales} in terms of the relevant dynamical timescales as a function of the dissipation strength. The time of maximum vortex clustering \(t(C_{\mathrm{max}})\) obtained from the numerical simulations is shown by filled points and compared to the characteristic vortex-depletion timescale \(\tau_{n_\mathrm{v}}\)\footnote{Although vortex decay exhibits different dynamical regimes, we extract $\tau_{n_\mathrm{v}}$ by fitting the vortex-number evolution with a stretched exponential $N_v(t)\sim\exp[-(t/\tau_{n_\mathrm{v}})^\alpha]$, treating both $\tau_{n_\mathrm{v}}$ and $\alpha$ as free parameters.}, the particle lifetime $\tau_\gamma$ and the crossover times \(t_s\) as extracted in Fig.~\ref{fig:evolution_energy_particles_velocity}(c). 
The timescale $\tau_{n_\mathrm{v}}$, extracted from the global decay of the vortex population, is mostly controlled by the early-stage depletion dynamics and provides an operational estimate of when vortex-mediated correlations can start to compete effectively with particle-loss-induced truncation. By contrast, the crossover times $t_s$, extracted from Fig.~\ref{fig:evolution_energy_particles_velocity}(c), identify the opposite boundary of the clustering-active window, namely when the remaining vortex population is no longer sufficient to sustain further efficient growth of correlations. 

Figure~\ref{fig:Cmax_timescales} therefore makes the physical picture explicit. For short particle lifetimes, clustering is particle-loss-limited because the condensate disappears before vortex correlations can fully develop. As the lifetime increases, the vortex sector has time to reorganize, and clustering grows. For even longer particle lifetimes, however, clustering becomes vortex-loss-limited: although the condensate survives longer, the buildup of order is no longer sustained once the dynamics cross the upper boundary \(t_s\). The strongest enhancement is found in the intermediate regime where the time of maximal clustering falls inside this finite dynamical window, bounded from below by \(\tau_{n_\mathrm{v}}\) and from above by \(t_s\).

The existence of an optimum beyond the conservative limit, together with the failure of maximal evaporative heating to predict maximal clustering, reflects a loss-induced modification of the finite dynamical window over which vortices reorganize. By contrast with pumped driven-dissipative condensates, where nonequilibrium gain can qualitatively modify vortex-pair interactions and even induce effective vortex-antivortex repulsion~\cite{Gladilin_2017, Gladilin_2019, wachtel2016electrodynamic}, the freely decaying system studied here contains only particle loss. The enhancement observed here is therefore not associated with gain-induced pair repulsion but with the compressible dynamics of the background fluid: varying the particle lifetime reshapes the redistribution among kinetic, interaction, compressible, and incompressible channels (see~\cite{SI} for a detailed discussion). 

In conclusion, dissipation enhances clustering by selecting a finite dynamical window in which vortex-mediated ordering can develop efficiently. 
Notably, the dissipation timescales associated with the optimal ordering regime fall within the experimentally tunable range of both cold atomic gases and fluids of light~\cite{SI}.
The lower edge of this window is set by the timescale on which vortex correlations begin to prevail over pure particle-loss truncation. At the same time, the upper edge is set by the loss of clustering efficiency once the residual vortex population becomes too depleted. Large-scale ordering is strongest when this window is widest, i.e., when background rarefaction and incompressible kinetic energy relaxation remain dynamically comparable over a finite time interval. 
The conservative limit is therefore not the regime of maximal ordering in our trapped system, and moderate dissipation can move the compressible quantum fluid away from the suboptimal conservative regime by selecting a finite interval in which vortex-mediated correlations can develop before the vortex population is depleted. 
The novelty of the present result is the identification of a dissipatively selected ordering regime of a compressible confined quantum fluid.

\

\paragraph{Acknowledgements.}
We are grateful for stimulating discussions with G. Martone,  N. P. Müller and M. Szymanska.
This work was financially supported within the ``NP Research, innovation and competitiveness for green and digital transition 2021-2027" (PN RIC 2021-2027) co-financed by the European Union, the Italian Ministry of Enterprises and Made in Italy (MIMIT) and the Italian Ministry of University and Research (MUR), through projects: AI-PHOQUS ``Artificial Intelligence and Advanced Networks Embedded in Photonics and Quantum Sciences and Technology", CUP B83C26000470007; ``Centro meridionale per l'innovazione quantistica", CUP: B49H26000300007; and ``Polo meridionale di innovazione per l'informazione quantistica", CUP: B42F26000460005. The authors acknowledge funding also from the Italian Ministry of University and Research (MUR) under the granting scheme FIS 3 (grant number FIS-2024-04047) and from the European Innovation Council (EIC) Pathfinder Programme in the frame of the European Union’s Horizon Europe initiative, through projects: ``Quantum Optical Networks based on Exciton-polaritons" (Q-ONE), Horizon-EIC-2022-Pathfinder Challenges, grant agreement No. 101115575 and ``Neuromorphic Polariton Accelerator" (PolArt), Horizon-EIC-2023-Pathfinder Open, grant agreement No. 101130304. Views and opinions expressed were however those of the authors only and do not necessarily reflect those of the European Union or EIC and SMEs Executive Agency (EISMEA). Neither the European Union nor the granting authority can be held responsible for them. The Center for Quantum-Enabled Computing project is carried out within the International Research Agendas programme of the Foundation for Polish Science co-financed by the European Union under the European Funds for Smart Economy 2021-2027 (FENG).
This study was conducted using the DARIAH HPC-AI cluster at CNR-NANOTEC in Lecce, funded by the ``MUR PON Ricerca e Innovazione 2014-2020" project, code PIR01$\_$00022 and H2IOSC Project - Humanities and cultural Heritage Italian Open Science Cloud funded by the European Union - NextGenerationEU - NRRP M4C2 - Project code IR0000029. 

%


\clearpage
\onecolumngrid
\beginsupplement


\makeatletter
\def\maketitle{
\@author@finish
\title@column\titleblock@produce
\suppressfloats[t]}
\makeatother

\title{Supplementary information for:\\  Dissipation-enhanced vortex clustering in a compressible quantum fluid}

\maketitle

\onecolumngrid
\beginsupplement

\section{Numerical simulations}

In this section, we focus on the numerical methods and details of the simulations studied in this work. Eq.~(1) of the main text is a dissipative Gross-Pitaevski Equation (dGPE), which describes the dynamics of the Bose field $\psi({\bf r},t)$ subject to linear losses $\propto \gamma$, and under a radial potential $V(\textbf{r})$ with diameter $D$.

We adopt parameters typical of experimental realization of polariton quantum fluids (see Ref.\cite{panico2023}): an effective mass $m = 0.22 \ \si{ps^2 meV \mu m^{-2}} = 3.52 \times 10^{-35} \ \si{kg}$, and an interaction strength $g = 5  \si{\mu eV \mu m^2}$. Eq.~(1) of the main text is numerically integrated using an eighth-order Runge-Kutta scheme on a $512^2$ spatial grid. The ratio $\alpha = D/L$ between the trap diameter $D=125 \mu m$ and the domain size $L= 312.5 \mu m$ is $\alpha = 0.4$. The initial state is chosen so to simulate a condensate with uniform density and where a vortex cloud is imprinted.
\begin{figure}[th!]
    \centering
    \includegraphics[width=.5\columnwidth]{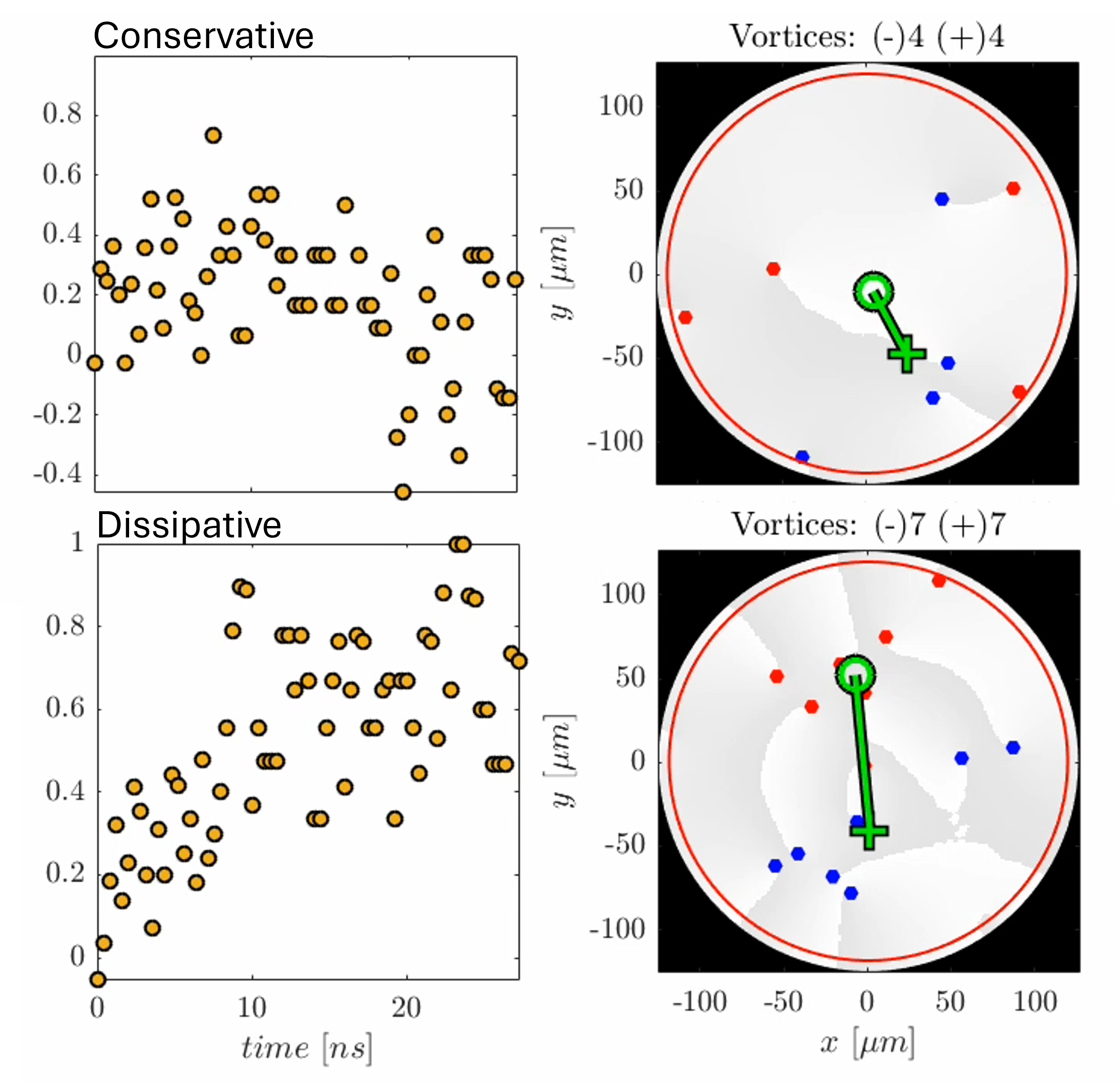}
    \caption{Dynamics of a single realization of the initial vortex distribution for conservative case (top row) and dissipative (bottom row) evolution with $\tau_\gamma = 20$ns. Left panels: clustering correlation function versus time. Right panels: condensate phase (grayscale background) together with vortex (blue points) and antivortex (red points) positions at a selected time $t=28$ns. While the clustering correlation decreases during conservative evolution, it increases toward unity in the dissipative case, demonstrating dissipation-enhanced vortex clustering.}
	\label{figSM:1}
\end{figure}
Each realisation of the numerical simulation corresponds to a random distribution of vortices. The position of the $i$-th vortex, centered at $\textbf{r}_i = (x_0-x_i,y_0-y_i)$, with $\textbf{r}_c =(x_0,y_0)$ the center of the potential, is randomly chosen at each stochastic realisation. \\For each realisation, a condensate is created with imprinted a vortex cloud. In order to do so, the wavefunction is initialized as $\psi(\textbf{r}) = \sqrt{n_0} \ \Pi_i [\textbf{r}_i/\sqrt{(\xi_0/\xi_c)^2 + \textbf{r}_i^2} \exp{(\pm i \theta_i)}]$, with $\xi_c$ a parameter regulating the shape of the vortex core.
\begin{figure*}[th!]
    \centering
    \includegraphics[width=0.7\textwidth]{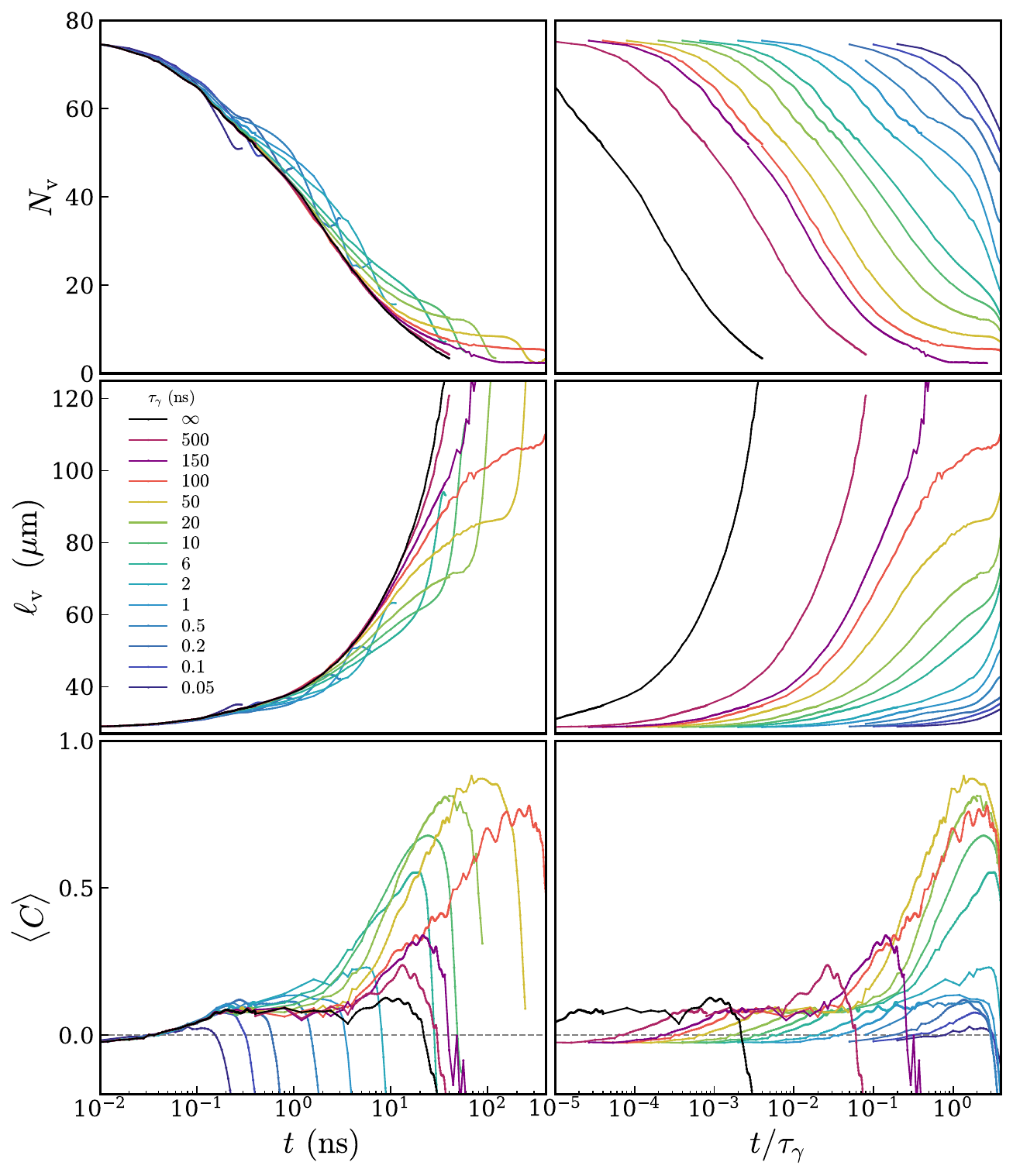}
    \caption{Number of vortices (top), mean intervortex distance (middle), and clustering correlation function (bottom) as a function of time for simulations with different dissipation timescales $\tau_\gamma$ (see labels). In the panels in the right column, time is made adimensional with the dissipative timescale $\tau_{\gamma}$.}
	\label{figSM:2}
\end{figure*}
During the temporal evolution, the vortex number and their positions are extracted from the phase gradients around closed paths of each grid point after a Gaussian filtering is applied in order to get rid of highest-order fluctuations~{(see Ref.~\cite{Comaron2021} for details on the vortex tracking algorithm)}.

In the numerical experiments, we use the disc diameter $D$ as a parameter controlling the initial inter-vortex length $\ell_\mathrm{v}^0 = D/\sqrt{N_\mathrm{v}^0}$, where $N_\mathrm{v}^0=80$ is the initial number of vortices. In doing so, we are careful to set $(\ell_\mathrm{v}^0/D) \le 0.1$, small enough to avoid finite-size effects. 

We chose a configuration of vortices that is random and free of clusters, by carefully checking that the initial condition for the dipole moment, 
averaged over the $\mathcal{N}$ realizations, always reads $\left< d \right>_\mathcal{N}(t=0)<3 \mathrm{\mu m}$. Moreover, vortex cores are separated by a distance $\ell^0_\mathrm{v}$ not smaller than $1.5 \mathrm{\mu m}$. Ensemble averaging is applied over $\mathcal{N}=5 \times 10^2$ different initial random vortex configurations. 

The dynamics of one numerical experiment for the conservative system, and for a dissipative system are shown in Fig.~\ref{figSM:1} (top and bottom rows, respectively). The figure illustrates a representative example of dissipation-enhanced vortex clustering. In the left panels, the clustering correlation function (yellow points) decreases over time in the conservative case, whereas it increases toward unity in the dissipative case. The right panels show the condensate phase (grayscale background) together with the vortex (blue points) and antivortex (red points) positions for the two dynamical regimes. To add on the numerical results presented in the main body, we show in Fig.~\ref{figSM:2} the ensemble-averaged evolution of the vortex number, the mean intervortex distance, and the clustering correlation function for all simulations performed for this study. Note, in the bottom row of this figure, that systems with $\tau_{\gamma} >100$ns are those for which clustering becomes vortex-loss-limited (see Fig.4 of the main body): this implies that the vortex clustering reaches its peak on a time scales shorter then the lifetime, $\tau_{\gamma}$.

\section{Analytical derivation of energy conservation laws for the GPE and dGPE.}

In this section, we derive the energy evolution law for thedissipative Gross--Pitaevskii equations. We show that the inclusion of a finite dissipation rate $\gamma$ leads to a monotonic decrease of the total energy $E(t)$, and that the decay rate is modified by the nonlinear interaction energy and therefore is not, in general, a simple exponential. Finally, we obtain analytical expressions for $E(t)$ and demonstrate that, as the interaction and potential energy contributions vanish at long times, the total energy approaches the kinetic energy and asymptotically decays exponentially with rate $\gamma$.

\subsection{Temporal evolution of total energy in the dissipative GPE}
\label{sec:dgpe_energy_decay}
The dissipative Gross--Pitaevskii equation we are considering is
\begin{equation}
\label{eq:S1}
i \hbar \frac{\partial \psi}{\partial t} = \left(
-\frac{\hbar^2}{2m} \nabla^2 + V(\mathbf{r}) + g |\psi|^2 - i \hbar \frac{\gamma}{2}
\right) \psi.
\end{equation}
For a time-independent potential, as we are considering, the total energy integral $E=E_{k} + E_{pot} + E_{int}$ is the sum of the three contributions of the kinetic, potential and interaction energies, respectively
\begin{equation}
\label{eq:S2}
E(t) = \int d\mathbf{r} \left[
\frac{\hbar^2}{2m} |\nabla \psi|^2 + V(\mathbf{r}) |\psi|^2 + \frac{g}{2} |\psi|^4
\right].
\end{equation}

We consider the temporal evolution of the total energy of the condensate,
\begin{equation}
\label{eq:S3}
\frac{dE(t)}{dt} = \int d\mathbf{r} \left[
\frac{\hbar^2}{2m} \frac{d}{dt} |\nabla \psi|^2 + V \frac{d}{dt} |\psi|^2 + \frac{g}{2} \frac{d}{dt} |\psi|^4
\right].
\end{equation}

Using the chain rule, we write
\begin{equation}
\label{eq:S4}
\frac{d}{dt} |\psi|^2 = \psi^* \frac{\partial \psi}{\partial t} + \psi \frac{\partial \psi^*}{\partial t},
\end{equation}
and
\begin{equation}
\label{eq:S5}
\frac{d}{dt} |\nabla \psi|^2 = \nabla \psi^* \cdot \nabla \frac{\partial \psi}{\partial t} + \nabla \psi \cdot \nabla \frac{\partial \psi^*}{\partial t},
\end{equation}
as well as
\begin{equation}
\label{eq:S6}
\frac{d}{dt} |\psi|^4 = 2 |\psi|^2 \frac{d}{dt} |\psi|^2.
\end{equation}

Therefore,
\begin{equation}
\label{eq:S7}
\begin{aligned}
\frac{dE}{dt} = \int d\mathbf{r} \Bigg[
& \frac{\hbar^2}{2m} \left( \nabla \psi^* \cdot \nabla \frac{\partial \psi}{\partial t} + \nabla \psi \cdot \nabla \frac{\partial \psi^*}{\partial t} \right) \\
& + V \left( \psi^* \frac{\partial \psi}{\partial t} + \psi \frac{\partial \psi^*}{\partial t} \right) \\
& + g |\psi|^2 \left( \psi^* \frac{\partial \psi}{\partial t} + \psi \frac{\partial \psi^*}{\partial t} \right)
\Bigg].
\end{aligned}
\end{equation}

Starting from eq.~\eqref{eq:S1}, we define the Hamiltonian operator ${\cal H}_{\infty}$ associated to the conservative evolution of the dynamics as ${\cal H}_{\infty}  = -\frac{\hbar^2}{2m} \nabla^2 + V + g |\psi|^2$. We can then re-write eq.~\eqref{eq:S1} as
\begin{equation}
\label{eq:S8}
\frac{\partial \psi}{\partial t} = -\frac{i}{\hbar} {\cal H}_{\infty} \psi - \frac{\gamma}{2} \psi,
\end{equation}

The complex conjugate is
\begin{equation}
\frac{\partial \psi^*}{\partial t} = \frac{i}{\hbar} {\cal H}_{\infty} \psi^* - \frac{\gamma}{2} \psi^*.
\end{equation}

Substitute these into the expression into eq.~\eqref{eq:S7}:
\begin{equation}
\label{eq:S9}
\begin{aligned}
\frac{dE}{dt} = \int d\mathbf{r} \Bigg[
& \frac{\hbar^2}{2m} \left( \nabla \psi^* \cdot \nabla \left( -\frac{i}{\hbar} {\cal H}_{\infty} \psi - \frac{\gamma}{2} \psi \right) + \nabla \psi \cdot \nabla \left( \frac{i}{\hbar} {\cal H}_{\infty} \psi^* - \frac{\gamma}{2} \psi^* \right) \right) \\
& + V \left( \psi^* \left( -\frac{i}{\hbar} {\cal H}_{\infty} \psi - \frac{\gamma}{2} \psi \right) + \psi \left( \frac{i}{\hbar} {\cal H}_{\infty} \psi^* - \frac{\gamma}{2} \psi^* \right) \right) \\
& + g |\psi|^2 \left( \psi^* \left( -\frac{i}{\hbar} {\cal H}_{\infty} \psi - \frac{\gamma}{2} \psi \right) + \psi \left( \frac{i}{\hbar} {\cal H}_{\infty} \psi^* - \frac{\gamma}{2} \psi^* \right)
\right)
\Bigg].
\end{aligned}
\end{equation}

We now terms: the imaginary parts cancel in the real part of the total energy, and the real terms are:
\begin{equation}
\label{eq:S10}
\frac{dE(t)}{dt} = -\gamma \int d\mathbf{r} \left[
\frac{\hbar^2}{2m} |\nabla \psi|^2 + V |\psi|^2 + g |\psi|^4
\right] = -\gamma D(t),
\end{equation}
where \( D(t) >0 \) is the instantaneous energy-like functional evaluated on the state \( \psi \). Note in the last term in the above equation does not coincide with the interaction energy, $E_{int} = \int d\mathbf{r} \frac{g}{2} |\psi|^4$, due to the prefactor.\\
In the dissipative Gross--Pitaevskii equation, the total energy decreases monotonically over time due to the non-Hermitian loss term. The decay rate is not proportional to the energy itself, but rather to the integral of the same functional form as the energy — evaluated on the instantaneous, non-stationary wavefunction. Thus:

\begin{equation}
\label{eq:S11}
\frac{dE(t)}{dt} = -\gamma D(t) < 0.
\end{equation}

As explicit above, this is not equal to \( -\gamma E(t) \), except in the noninteracting limit \( g \to 0 \). The nonlinear term causes an additional contribution to the decay rate of the energy.

Alternatively, we can also write more explicitly
\begin{equation}
\label{eq:S12}
\frac{dE(t)}{dt} = -\gamma E(t) - \frac{\gamma g}{2} \int d\mathbf{r} |\psi|^4= -\gamma E(t) - \gamma E_{\mathrm{int}}(t).
\end{equation}

\subsection{Analytical solution of $E(t)$}
\label{sec:analytical_energy_solution}

We consider the following solutions of eq.~\eqref{eq:S12}

\medskip

\textbf{Case 1:} \( E_{\mathrm{int}}(t) \) varies on a time scale much larger than that of kinetic energy, such that we can write in a first approximation \( E_{\mathrm{int}}(t) = E_{\mathrm{int}} = \mathrm{cnst} \).

The equation becomes
\begin{equation}
\label{eq:S13}
\frac{dE}{dt} = -\gamma E(t) - \gamma E_{\mathrm{int}}.
\end{equation}

This is a linear first-order ODE with solution found via the integrating factor method. Fixing the initial condition $E(t=0)=E_0$, the solution is simply
\begin{equation}
\label{eq:S14}
E(t) = \left( E_0 + E_{\mathrm{int}} \right) e^{-\gamma t} - E_{\mathrm{int}}.
\end{equation}

\vspace{1em}

\textbf{Case 2:} \( E_{\mathrm{int}}(t) \) is time-dependent, and again the initial condition $E(t=0)=E_0$. 

Multiply the original equation by \( e^{\gamma t} \):
\begin{equation}
\label{eq:S15}
e^{\gamma t} \frac{dE}{dt} + \gamma e^{\gamma t} E = - \gamma e^{\gamma t} E_{\mathrm{int}}(t),
\end{equation}
which implies
\begin{equation}
\label{eq:S16}
\frac{d}{dt} \left( e^{\gamma t} E(t) \right) = -\gamma e^{\gamma t} E_{\mathrm{int}}(t).
\end{equation}

Integrate from 0 to \( t \):
\begin{equation}
\label{eq:S17}
e^{\gamma t} E(t) - E_0 = -\gamma \int_0^t e^{\gamma s} E_{\mathrm{int}}(s) \, ds,
\end{equation}
and solve for \( E(t) \):
\begin{equation}
\label{eq:S18}
{
E(t) = e^{-\gamma t} \left[ E_0 - \gamma \int_0^t e^{\gamma s} E_{\mathrm{int}}(s) \, ds \right].
}
\end{equation}

The analytical solution highlights the role of the interaction energy in determining the energy relaxation dynamics. When the interaction contribution $E_{\mathrm{int}}(t)$ decreases with time, the total energy $E(t)$ relaxes more rapidly than a purely exponential decay. In the limiting case of a time-independent interaction energy, $E_{\mathrm{int}}(t)=E_{\mathrm{int}}$, the solution reduces to an inhomogeneous exponential relaxation. Conversely, if the interaction energy vanishes asymptotically,
$$
E_{\mathrm{int}}(t)\rightarrow 0 \qquad (t\rightarrow\infty),$$

the nonlinear correction becomes negligible and the total energy approaches at long times the simple exponential form $E(t)\sim E_0\,e^{-\gamma t}\,.$
Therefore, deviations from exponential relaxation are a direct consequence of the nonlinear interaction energy, while the long-time dynamics recovers the exponential decay expected for an effectively non-interacting system.

\section{{Energy decomposition and redistribution}}
\begin{figure*}[t]
    \centering
    \includegraphics[width=\textwidth]{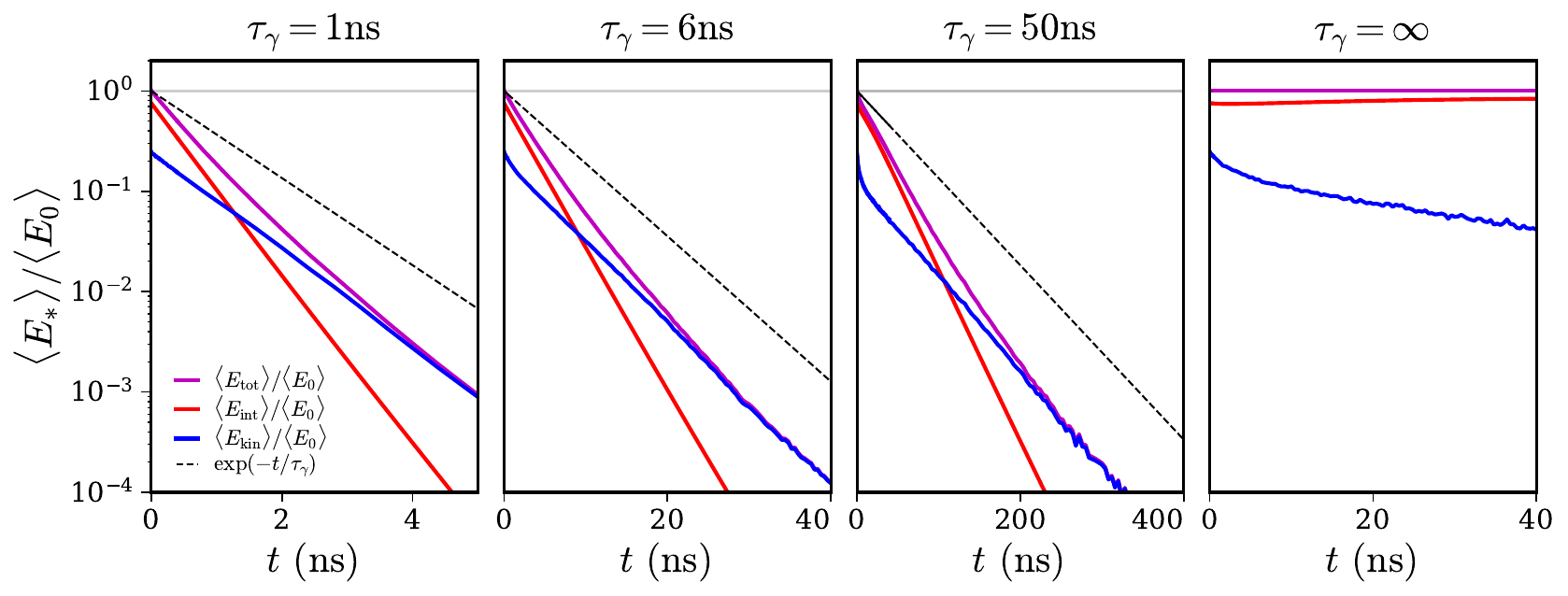}
    \caption{Lin-log plot of the temporal evolution for $t>1$ns of the total (magenta), interaction (red), and kinetic (blue) energies, normalized to the initial total energy, ensemble averaged, for three different dissipative cases and the conservative one. The exponential decay law $e^{-t/\tau_\gamma}$ (black dashed line), is also shown.}
	\label{figSM:3}
\end{figure*}

We analyze the temporal evolution of the different terms in the total energy budget (\ref{eq:S2}). In Fig.~\ref{figSM:3}, the total energy, together with the kinetic and interaction ones are plotted, for different particle lifetimes. The potential energy being very small is not shown. Next, by applying the Helmholtz decomposition to the density-weighted superfluid velocity $\textbf{u}(\textbf{r},t) = \sqrt{n(\textbf{r},t)} \textbf{v}(\textbf{r},t)$, with $\textbf{v}(\textbf{r},t) = \frac{\hbar}{m} \nabla\theta(\textbf{r},t)$ the velocity of the superfluid, we separate the compressible ($E_\mathrm{ck}$) and incompressible ($E_\mathrm{ik}$) components of the kinetic energy, which can be attributed to the sound and vortex excitations of the fluid, respectively (see also SI in \cite{panico2023}).\\
The Helmholtz decomposition highlights a key feature of our set-up: for all cases, the kinetic energy is dominated by its compressible component, which accounts for approximately 94\% of the total kinetic energy over the whole evolution, although small fluctuations are present.
This suggests that the system’s dynamics are primarily governed by density fluctuations rather than vortex-induced flow. However, while vortices themselves carry little energy, their annihilation triggers localized perturbations. These waves interfere and accumulate, generating sharp density modulations. Since the interaction energy scales with the fourth power of the density amplitude, $|\psi|^4$, these modulations can significantly enhance the interaction energy, effectively transferring energy from coherent flow into pressure-like peaks within the condensate.

In the conservative case, Fig.~\ref{figSM:3}(right), as expected, the total energy $E_\mathrm{tot}$ remains conserved, while the kinetic energy decreases. Simultaneously, we observe a corresponding increase in the interaction energy, indicating a transfer of energy from kinetic to interaction components. This behavior can be understood in terms of energy redistribution during the system's evolution. As the condensate evolves, vortex–antivortex pairs annihilate, and individual vortices can migrate toward the boundary of the trap and exit the system. \\
\begin{figure*}[t]
	\centering
       \includegraphics[width=\linewidth]{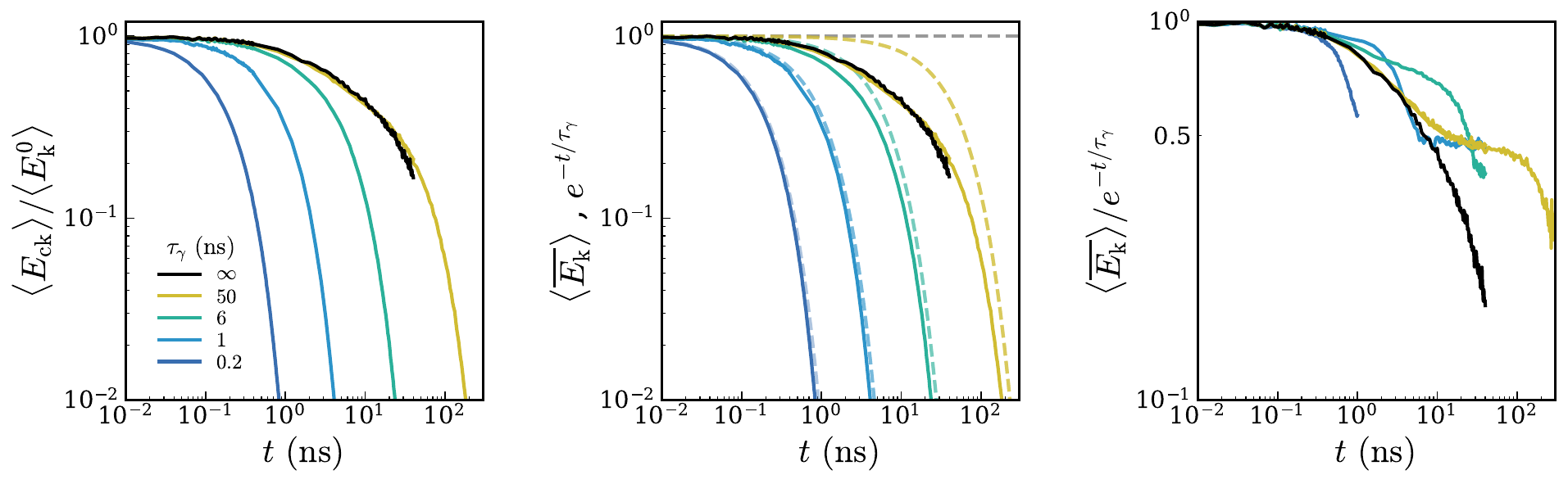}
	\caption{(Left panel) Time evolution of the ensemble-averaged compressible part of the kinetic energy $\langle E_{ck}\rangle$ normalised to the total kinetic energy at initial time $\langle E_{k}^0\rangle$, for a few dissipative cases compared to the conservative one. (Central panel) The time decay of the normalised total kinetic energy $\langle {\overline E_{k}}\rangle$, where 
    $\langle {\overline E_{k}}\rangle\equiv \langle E_{k}(t)\rangle/\langle E_{k}^0\rangle$ (full lines); the exponential decay functions $e^{-t/\tau_{\gamma}}$ for various $\tau_{\gamma}$ (dashed lines), the grey dashed line refers to the conservative case. (Right panel) The time decay of the (normalised) total kinetic energy per particle $ \langle \overline{E_{k}}\rangle/(e^{-t/\tau_{\gamma}})$ for the dissipative cases; for the conservative cases the black curve is just $\langle {\overline E_{k}}\rangle$.}
	\label{figSM:4}
\end{figure*}
In the dissipative cases (first three panels) of Fig.~\ref{figSM:3}), we show that dissipation leads to a monotonic decay of both particle density and energy, reflecting the non-Hermitian nature of the system.
The total energy generally decays over time due to losses, but it does not necessarily follow the same decay rate as the particle density $n(t)/n(0) \simeq \exp(-t/\tau)$ (black dashed lines in Fig.~\ref{figSM:3}). Indeed, the energy decay depends on the evolving balance of kinetic, potential, and interaction terms.
As previously discussed, for a time dependent interaction energy $E_\mathrm{int}(t)$, we have that 
$$
E(t) = e^{-\gamma t} \left[ E_0 - \gamma \int_0^t e^{\gamma s} E_{\mathrm{int}}(s) \, ds \right],
$$
showing that the total energy $ E(t)$ decays faster than a simple exponential and recovers the exponential decay only at long times. This is in agreement with the observations in Figs.~\ref{figSM:3}. 

We observe a striking difference between the conservative and dissipative cases: in the latter cases, we observe a crossover from an interaction-dominated regime ($E_{\mathrm{int}}$) at early times to a kinetic-energy-dominated regime ($E_{\mathrm{kin}}$) at late times. This transition generally occurs close to the peak of the clustering correlation (not shown). This suggests a connection between the interplay of interactions and kinetic energy, and the decay of clustering. 

Finally, in Fig.~\ref{figSM:4}, we plot the time evolution of the compressible and total kinetic energies of the conservative case, compared to a few dissipative cases. In the left panel, we appreciate that the compressible kinetic energy is initially almost equal to the total kinetic energy, and remains so for the whole temporal evolution (not shown). \\In Fig.~\ref{figSM:4}(central panel), we observe that for dissipative systems with short lifetimes (e.g., $\tau_{\mathrm{\gamma}} = 0.2\,\mathrm{ns}$), the decay of the total kinetic energy $E_{k}$ closely follows the particle density exponential decay, indicating that the particle loss halts the whole dynamics. In this regime, vortices do not have sufficient time to interact and develop statistical nonlinear behavior.\\
As $\tau_{\mathrm{\gamma}}$ increases (e.g., $\tau_{\mathrm{\gamma}} = 1\,\mathrm{ns}$, $6\,\mathrm{ns}$), the decay of the kinetic energy progressively accelerates with respect to the density. This behavior is linked to the afore-mentioned mechanism where kinetic energy is dissipated through vortex-pairs annihilation and generally vortex loss, responsible for the onset of clustering processes. At increasing $\tau_{\gamma}$, we hence observe - at fixed time-  that in condensates with larger $\tau_{\gamma}$ the kinetic energy per particle decays faster, as shown in Fig.~\ref{figSM:4}(right panel). For even longer lifetimes (e.g., $\tau_{\mathrm{\gamma}} = 50\,\mathrm{ns}$), the kinetic energy initially saturates to a value comparable with that of the conservative case, before eventually recovering an exponential decay. This is consistent with the earlier discussion: at sufficiently long times, kinetic energy becomes the dominant contribution to the total energy, which asymptotically decays exponentially. Strikingly, Fig.~\ref{figSM:4}(right panel) shows that this crossover corresponds to a flattening of the mean kinetic energy per particle.

\

\subsection{Effective damping parameter}
In this paragraph we comment on how the timescales and the model investigated in this work relates to Bose gases in general.
Starting from the dissipative GPE with linear losses,
\begin{equation}
i\hbar\frac{\partial\psi}{\partial t}=\left[{\cal H}_{\infty} -\frac{i\hbar}{2}\gamma\right]\psi,
\qquad \gamma=\tau_\gamma^{-1},
\end{equation}
one can define an \emph{effective} dimensionless damping parameter comparable to the atomic dGPE form
\begin{equation}
i\hbar\frac{\partial\psi}{\partial t}=(1-i\Gamma)\,{\cal H}_{\infty}\psi,
\end{equation}
by matching the imaginary terms assuming the field evolves at a typical energy scale ${\cal H}_{\infty}\psi\simeq \mu\psi$ (with $\mu\simeq gn$). This gives
\begin{equation}
\Gamma \simeq \frac{\hbar\gamma}{2\mu}
=\frac{\hbar}{2\mu\,\tau_{\gamma}}.
\end{equation}
With our typical parameters ($n\simeq 400~\mu{\rm m}^{-2}$, $g\simeq 5\times10^{-3}~{\rm meV}\,\mu{\rm m}^2\Rightarrow \mu\simeq 2~{\rm meV}$), lifetimes in the range $\tau_{\gamma}\sim 10$--$100$~ns correspond to very small $\Gamma\sim 10^{-5}$--$10^{-6}$, i.e.\ weak damping on the intrinsic nonlinear timescale.
These ranges of parameters correspond to typical cold-atoms experimental ranges for $\Gamma$ \cite{weiler2008spontaneous,Rooney2013,Comaron2019}.
However, in atomic condensates the dissipative term corresponds to energy relaxation and mutual-friction-like damping, which can additionally slow vortex motion beyond the effect of particle loss alone.


\begin{thebibliography}{44}%
\makeatletter
\providecommand \@ifxundefined [1]{%
 \@ifx{#1\undefined}
}%
\providecommand \@ifnum [1]{%
 \ifnum #1\expandafter \@firstoftwo
 \else \expandafter \@secondoftwo
 \fi
}%
\providecommand \@ifx [1]{%
 \ifx #1\expandafter \@firstoftwo
 \else \expandafter \@secondoftwo
 \fi
}%
\providecommand \natexlab [1]{#1}%
\providecommand \enquote  [1]{``#1''}%
\providecommand \bibnamefont  [1]{#1}%
\providecommand \bibfnamefont [1]{#1}%
\providecommand \citenamefont [1]{#1}%
\providecommand \href@noop [0]{\@secondoftwo}%
\providecommand \href [0]{\begingroup \@sanitize@url \@href}%
\providecommand \@href[1]{\@@startlink{#1}\@@href}%
\providecommand \@@href[1]{\endgroup#1\@@endlink}%
\providecommand \@sanitize@url [0]{\catcode `\\12\catcode `\$12\catcode `\&12\catcode `\#12\catcode `\^12\catcode `\_12\catcode `\%12\relax}%
\providecommand \@@startlink[1]{}%
\providecommand \@@endlink[0]{}%
\providecommand \url  [0]{\begingroup\@sanitize@url \@url }%
\providecommand \@url [1]{\endgroup\@href {#1}{\urlprefix }}%
\providecommand \urlprefix  [0]{URL }%
\providecommand \Eprint [0]{\href }%
\providecommand \doibase [0]{https://doi.org/}%
\providecommand \selectlanguage [0]{\@gobble}%
\providecommand \bibinfo  [0]{\@secondoftwo}%
\providecommand \bibfield  [0]{\@secondoftwo}%
\providecommand \translation [1]{[#1]}%
\providecommand \BibitemOpen [0]{}%
\providecommand \bibitemStop [0]{}%
\providecommand \bibitemNoStop [0]{.\EOS\space}%
\providecommand \EOS [0]{\spacefactor3000\relax}%
\providecommand \BibitemShut  [1]{\csname bibitem#1\endcsname}%
\let\auto@bib@innerbib\@empty
\bibitem [{\citenamefont {Frisch}(1995)}]{Frisch1995}%
  \BibitemOpen
  \bibfield  {author} {\bibinfo {author} {\bibfnamefont {U.}~\bibnamefont {Frisch}},\ }\href@noop {} {\emph {\bibinfo {title} {Turbulence: The Legacy of A. N. Kolmogorov}}}\ (\bibinfo  {publisher} {Cambridge University Press},\ \bibinfo {year} {1995})\BibitemShut {NoStop}%
\bibitem [{\citenamefont {Kraichnan}(1967)}]{kraichnan1967inertial}%
  \BibitemOpen
  \bibfield  {author} {\bibinfo {author} {\bibfnamefont {R.~H.}\ \bibnamefont {Kraichnan}},\ }\bibfield  {title} {\bibinfo {title} {Inertial ranges in two-dimensional turbulence},\ }\href {https://doi.org/10.1063/1.1762301} {\bibfield  {journal} {\bibinfo  {journal} {The Physics of Fluids}\ }\textbf {\bibinfo {volume} {10}},\ \bibinfo {pages} {1417} (\bibinfo {year} {1967})}\BibitemShut {NoStop}%
\bibitem [{\citenamefont {Onsager}(1949)}]{onsager1949statistical}%
  \BibitemOpen
  \bibfield  {author} {\bibinfo {author} {\bibfnamefont {L.}~\bibnamefont {Onsager}},\ }\bibfield  {title} {\bibinfo {title} {Statistical hydrodynamics},\ }\href {https://link.springer.com/article/10.1007/BF02780991} {\bibfield  {journal} {\bibinfo  {journal} {Il Nuovo Cimento (1943-1954)}\ }\textbf {\bibinfo {volume} {6}},\ \bibinfo {pages} {279} (\bibinfo {year} {1949})}\BibitemShut {NoStop}%
\bibitem [{\citenamefont {Boffetta}\ and\ \citenamefont {Ecke}(2012)}]{BEreview}%
  \BibitemOpen
  \bibfield  {author} {\bibinfo {author} {\bibfnamefont {G.}~\bibnamefont {Boffetta}}\ and\ \bibinfo {author} {\bibfnamefont {R.~E.}\ \bibnamefont {Ecke}},\ }\bibfield  {title} {\bibinfo {title} {Two-dimensional turbulence},\ }\href {https://doi.org/10.1146/annurev-fluid-120710-101240} {\bibfield  {journal} {\bibinfo  {journal} {Ann. Rev. Fluid Mech.}\ }\textbf {\bibinfo {volume} {44}},\ \bibinfo {pages} {427} (\bibinfo {year} {2012})}\BibitemShut {NoStop}%
\bibitem [{\citenamefont {Groszek}\ \emph {et~al.}(2018)\citenamefont {Groszek}, \citenamefont {Davis}, \citenamefont {Paganin}, \citenamefont {Helmerson},\ and\ \citenamefont {Simula}}]{Groszek2018}%
  \BibitemOpen
  \bibfield  {author} {\bibinfo {author} {\bibfnamefont {A.~J.}\ \bibnamefont {Groszek}}, \bibinfo {author} {\bibfnamefont {M.~J.}\ \bibnamefont {Davis}}, \bibinfo {author} {\bibfnamefont {D.~M.}\ \bibnamefont {Paganin}}, \bibinfo {author} {\bibfnamefont {K.}~\bibnamefont {Helmerson}},\ and\ \bibinfo {author} {\bibfnamefont {T.~P.}\ \bibnamefont {Simula}},\ }\bibfield  {title} {\bibinfo {title} {Vortex thermometry for turbulent two-dimensional fluids},\ }\href {https://doi.org/10.1103/PhysRevLett.120.034504} {\bibfield  {journal} {\bibinfo  {journal} {Phys. Rev. Lett.}\ }\textbf {\bibinfo {volume} {120}},\ \bibinfo {pages} {034504} (\bibinfo {year} {2018})}\BibitemShut {NoStop}%
\bibitem [{\citenamefont {Simula}\ \emph {et~al.}(2014)\citenamefont {Simula}, \citenamefont {Davis},\ and\ \citenamefont {Helmerson}}]{simula2014prl}%
  \BibitemOpen
  \bibfield  {author} {\bibinfo {author} {\bibfnamefont {T.}~\bibnamefont {Simula}}, \bibinfo {author} {\bibfnamefont {M.~J.}\ \bibnamefont {Davis}},\ and\ \bibinfo {author} {\bibfnamefont {K.}~\bibnamefont {Helmerson}},\ }\bibfield  {title} {\bibinfo {title} {Emergence of order from turbulence in an isolated planar superfluid},\ }\href {https://doi.org/10.1103/PhysRevLett.113.165302} {\bibfield  {journal} {\bibinfo  {journal} {Phys. Rev. Lett.}\ }\textbf {\bibinfo {volume} {113}},\ \bibinfo {pages} {165302} (\bibinfo {year} {2014})}\BibitemShut {NoStop}%
\bibitem [{\citenamefont {Johnstone}\ \emph {et~al.}(2019)\citenamefont {Johnstone}, \citenamefont {Groszek}, \citenamefont {Starkey}, \citenamefont {Billington}, \citenamefont {Simula},\ and\ \citenamefont {Helmerson}}]{johnstone2019evolution}%
  \BibitemOpen
  \bibfield  {author} {\bibinfo {author} {\bibfnamefont {S.~P.}\ \bibnamefont {Johnstone}}, \bibinfo {author} {\bibfnamefont {A.~J.}\ \bibnamefont {Groszek}}, \bibinfo {author} {\bibfnamefont {P.~T.}\ \bibnamefont {Starkey}}, \bibinfo {author} {\bibfnamefont {C.~J.}\ \bibnamefont {Billington}}, \bibinfo {author} {\bibfnamefont {T.~P.}\ \bibnamefont {Simula}},\ and\ \bibinfo {author} {\bibfnamefont {K.}~\bibnamefont {Helmerson}},\ }\bibfield  {title} {\bibinfo {title} {Evolution of large-scale flow from turbulence in a two-dimensional superfluid},\ }\href {https://www.science.org/doi/10.1126/science.aat5793} {\bibfield  {journal} {\bibinfo  {journal} {Science}\ }\textbf {\bibinfo {volume} {364}},\ \bibinfo {pages} {1267} (\bibinfo {year} {2019})}\BibitemShut {NoStop}%
\bibitem [{\citenamefont {Gauthier}\ \emph {et~al.}(2019)\citenamefont {Gauthier}, \citenamefont {Reeves}, \citenamefont {Yu}, \citenamefont {Bradley}, \citenamefont {Baker}, \citenamefont {Bell}, \citenamefont {Rubinsztein-Dunlop}, \citenamefont {Davis},\ and\ \citenamefont {Neely}}]{gauthier2019giant}%
  \BibitemOpen
  \bibfield  {author} {\bibinfo {author} {\bibfnamefont {G.}~\bibnamefont {Gauthier}}, \bibinfo {author} {\bibfnamefont {M.~T.}\ \bibnamefont {Reeves}}, \bibinfo {author} {\bibfnamefont {X.}~\bibnamefont {Yu}}, \bibinfo {author} {\bibfnamefont {A.~S.}\ \bibnamefont {Bradley}}, \bibinfo {author} {\bibfnamefont {M.~A.}\ \bibnamefont {Baker}}, \bibinfo {author} {\bibfnamefont {T.~A.}\ \bibnamefont {Bell}}, \bibinfo {author} {\bibfnamefont {H.}~\bibnamefont {Rubinsztein-Dunlop}}, \bibinfo {author} {\bibfnamefont {M.~J.}\ \bibnamefont {Davis}},\ and\ \bibinfo {author} {\bibfnamefont {T.~W.}\ \bibnamefont {Neely}},\ }\bibfield  {title} {\bibinfo {title} {Giant vortex clusters in a two-dimensional quantum fluid},\ }\href {https://www.science.org/doi/10.1126/science.aat5718} {\bibfield  {journal} {\bibinfo  {journal} {Science}\ }\textbf {\bibinfo {volume} {364}},\ \bibinfo {pages} {1264} (\bibinfo {year} {2019})}\BibitemShut {NoStop}%
\bibitem [{\citenamefont {Berloff}(2010)}]{Berloff2010Turbulence}%
  \BibitemOpen
  \bibfield  {author} {\bibinfo {author} {\bibfnamefont {N.~G.}\ \bibnamefont {Berloff}},\ }\bibfield  {title} {\bibinfo {title} {Turbulence in exciton-polariton condensates},\ }\href@noop {} {\  (\bibinfo {year} {2010})},\ \bibinfo {note} {arXiv:1010.5225},\ \Eprint {https://arxiv.org/abs/1010.5225} {arXiv:1010.5225 [cond-mat.quant-gas]} \BibitemShut {NoStop}%
\bibitem [{\citenamefont {Koniakhin}\ \emph {et~al.}(2020)\citenamefont {Koniakhin}, \citenamefont {Bleu}, \citenamefont {Malpuech},\ and\ \citenamefont {Solnyshkov}}]{Koniakhin2020CSF}%
  \BibitemOpen
  \bibfield  {author} {\bibinfo {author} {\bibfnamefont {S.~V.}\ \bibnamefont {Koniakhin}}, \bibinfo {author} {\bibfnamefont {O.}~\bibnamefont {Bleu}}, \bibinfo {author} {\bibfnamefont {G.}~\bibnamefont {Malpuech}},\ and\ \bibinfo {author} {\bibfnamefont {D.~D.}\ \bibnamefont {Solnyshkov}},\ }\bibfield  {title} {\bibinfo {title} {2d quantum turbulence in a polariton quantum fluid},\ }\href {https://doi.org/10.1016/j.chaos.2019.109574} {\bibfield  {journal} {\bibinfo  {journal} {Chaos, Solitons \& Fractals}\ }\textbf {\bibinfo {volume} {132}},\ \bibinfo {pages} {109574} (\bibinfo {year} {2020})}\BibitemShut {NoStop}%
\bibitem [{\citenamefont {Panico}\ \emph {et~al.}(2023{\natexlab{a}})\citenamefont {Panico}, \citenamefont {Comaron}, \citenamefont {Matuszewski}, \citenamefont {Lanotte}, \citenamefont {Trypogeorgos}, \citenamefont {Gigli}, \citenamefont {Giorgi}, \citenamefont {Ardizzone}, \citenamefont {Sanvitto},\ and\ \citenamefont {Ballarini}}]{panico2023}%
  \BibitemOpen
  \bibfield  {author} {\bibinfo {author} {\bibfnamefont {R.}~\bibnamefont {Panico}}, \bibinfo {author} {\bibfnamefont {P.}~\bibnamefont {Comaron}}, \bibinfo {author} {\bibfnamefont {M.}~\bibnamefont {Matuszewski}}, \bibinfo {author} {\bibfnamefont {A.~S.}\ \bibnamefont {Lanotte}}, \bibinfo {author} {\bibfnamefont {D.}~\bibnamefont {Trypogeorgos}}, \bibinfo {author} {\bibfnamefont {G.}~\bibnamefont {Gigli}}, \bibinfo {author} {\bibfnamefont {M.~D.}\ \bibnamefont {Giorgi}}, \bibinfo {author} {\bibfnamefont {V.}~\bibnamefont {Ardizzone}}, \bibinfo {author} {\bibfnamefont {D.}~\bibnamefont {Sanvitto}},\ and\ \bibinfo {author} {\bibfnamefont {D.}~\bibnamefont {Ballarini}},\ }\bibfield  {title} {\bibinfo {title} {Onset of vortex clustering and inverse energy cascade in dissipative quantum fluids},\ }\href {https://doi.org/10.1038/s41566-023-01174-4} {\bibfield  {journal} {\bibinfo  {journal} {Nature Photonics}\ }\textbf {\bibinfo {volume} {17}},\ \bibinfo {pages} {451} (\bibinfo {year}
  {2023}{\natexlab{a}})}\BibitemShut {NoStop}%
\bibitem [{\citenamefont {Depaepe}\ \emph {et~al.}(2026)\citenamefont {Depaepe}, \citenamefont {Ouahrouche}, \citenamefont {Amo},\ and\ \citenamefont {Hainaut}}]{Depaepe2026Counterflow}%
  \BibitemOpen
  \bibfield  {author} {\bibinfo {author} {\bibfnamefont {L.}~\bibnamefont {Depaepe}}, \bibinfo {author} {\bibfnamefont {K.}~\bibnamefont {Ouahrouche}}, \bibinfo {author} {\bibfnamefont {A.}~\bibnamefont {Amo}},\ and\ \bibinfo {author} {\bibfnamefont {C.}~\bibnamefont {Hainaut}},\ }\bibfield  {title} {\bibinfo {title} {Emergence of turbulence in a counterflow geometry of 2d polariton quantum fluids},\ }\href@noop {} {\  (\bibinfo {year} {2026})},\ \Eprint {https://arxiv.org/abs/2603.05125} {arXiv:2603.05125} \BibitemShut {NoStop}%
\bibitem [{\citenamefont {Baker-Rasooli}\ \emph {et~al.}(2023)\citenamefont {Baker-Rasooli}, \citenamefont {Liu}, \citenamefont {Aladjidi}, \citenamefont {Bramati},\ and\ \citenamefont {Glorieux}}]{BakerRasooli2023PRA}%
  \BibitemOpen
  \bibfield  {author} {\bibinfo {author} {\bibfnamefont {M.}~\bibnamefont {Baker-Rasooli}}, \bibinfo {author} {\bibfnamefont {W.}~\bibnamefont {Liu}}, \bibinfo {author} {\bibfnamefont {T.}~\bibnamefont {Aladjidi}}, \bibinfo {author} {\bibfnamefont {A.}~\bibnamefont {Bramati}},\ and\ \bibinfo {author} {\bibfnamefont {Q.}~\bibnamefont {Glorieux}},\ }\bibfield  {title} {\bibinfo {title} {Turbulent dynamics in a two-dimensional paraxial fluid of light},\ }\href {https://doi.org/10.1103/PhysRevA.108.063512} {\bibfield  {journal} {\bibinfo  {journal} {Phys. Rev. A}\ }\textbf {\bibinfo {volume} {108}},\ \bibinfo {pages} {063512} (\bibinfo {year} {2023})}\BibitemShut {NoStop}%
\bibitem [{\citenamefont {Valani}\ \emph {et~al.}(2018)\citenamefont {Valani}, \citenamefont {Groszek},\ and\ \citenamefont {Simula}}]{Valani_2018}%
  \BibitemOpen
  \bibfield  {author} {\bibinfo {author} {\bibfnamefont {R.~N.}\ \bibnamefont {Valani}}, \bibinfo {author} {\bibfnamefont {A.~J.}\ \bibnamefont {Groszek}},\ and\ \bibinfo {author} {\bibfnamefont {T.~P.}\ \bibnamefont {Simula}},\ }\bibfield  {title} {\bibinfo {title} {Einstein{\textendash}bose condensation of onsager vortices},\ }\href@noop {} {\bibfield  {journal} {\bibinfo  {journal} {New Journal of Physics}\ }\textbf {\bibinfo {volume} {20}},\ \bibinfo {pages} {053038} (\bibinfo {year} {2018})}\BibitemShut {NoStop}%
\bibitem [{\citenamefont {Billam}\ \emph {et~al.}(2014)\citenamefont {Billam}, \citenamefont {Reeves}, \citenamefont {Anderson},\ and\ \citenamefont {Bradley}}]{Billam2014}%
  \BibitemOpen
  \bibfield  {author} {\bibinfo {author} {\bibfnamefont {T.~P.}\ \bibnamefont {Billam}}, \bibinfo {author} {\bibfnamefont {M.~T.}\ \bibnamefont {Reeves}}, \bibinfo {author} {\bibfnamefont {B.~P.}\ \bibnamefont {Anderson}},\ and\ \bibinfo {author} {\bibfnamefont {A.~S.}\ \bibnamefont {Bradley}},\ }\bibfield  {title} {\bibinfo {title} {Onsager-kraichnan condensation in decaying two-dimensional quantum turbulence},\ }\href {https://doi.org/10.1103/PhysRevLett.112.145301} {\bibfield  {journal} {\bibinfo  {journal} {Phys. Rev. Lett.}\ }\textbf {\bibinfo {volume} {112}},\ \bibinfo {pages} {145301} (\bibinfo {year} {2014})}\BibitemShut {NoStop}%
\bibitem [{\citenamefont {Skipp}\ \emph {et~al.}(2023)\citenamefont {Skipp}, \citenamefont {Laurie},\ and\ \citenamefont {Nazarenko}}]{Skipp2023}%
  \BibitemOpen
  \bibfield  {author} {\bibinfo {author} {\bibfnamefont {J.}~\bibnamefont {Skipp}}, \bibinfo {author} {\bibfnamefont {J.}~\bibnamefont {Laurie}},\ and\ \bibinfo {author} {\bibfnamefont {S.}~\bibnamefont {Nazarenko}},\ }\bibfield  {title} {\bibinfo {title} {Hamiltonian derivation of the point vortex model from the two-dimensional nonlinear schr\"odinger equation},\ }\href {https://doi.org/10.1103/PhysRevE.107.025107} {\bibfield  {journal} {\bibinfo  {journal} {Phys. Rev. E}\ }\textbf {\bibinfo {volume} {107}},\ \bibinfo {pages} {025107} (\bibinfo {year} {2023})}\BibitemShut {NoStop}%
\bibitem [{\citenamefont {Tattersall}\ \emph {et~al.}(2025)\citenamefont {Tattersall}, \citenamefont {Baggaley},\ and\ \citenamefont {Billam}}]{tattersall2025out}%
  \BibitemOpen
  \bibfield  {author} {\bibinfo {author} {\bibfnamefont {R.~J.}\ \bibnamefont {Tattersall}}, \bibinfo {author} {\bibfnamefont {A.~W.}\ \bibnamefont {Baggaley}},\ and\ \bibinfo {author} {\bibfnamefont {T.~P.}\ \bibnamefont {Billam}},\ }\bibfield  {title} {\bibinfo {title} {Out-of-equilibrium behavior of quantum vortices: A comparison of point vortex dynamics and fokker-planck evolution},\ }\href {https://doi.org/10.1103/l8yk-kk5c} {\bibfield  {journal} {\bibinfo  {journal} {Phys. Rev. A}\ }\textbf {\bibinfo {volume} {112}},\ \bibinfo {pages} {013313} (\bibinfo {year} {2025})}\BibitemShut {NoStop}%
\bibitem [{\citenamefont {Billam}\ \emph {et~al.}(2015)\citenamefont {Billam}, \citenamefont {Reeves},\ and\ \citenamefont {Bradley}}]{Billam2015}%
  \BibitemOpen
  \bibfield  {author} {\bibinfo {author} {\bibfnamefont {T.~P.}\ \bibnamefont {Billam}}, \bibinfo {author} {\bibfnamefont {M.~T.}\ \bibnamefont {Reeves}},\ and\ \bibinfo {author} {\bibfnamefont {A.~S.}\ \bibnamefont {Bradley}},\ }\bibfield  {title} {\bibinfo {title} {Spectral energy transport in two-dimensional quantum vortex dynamics},\ }\href {https://doi.org/10.1103/PhysRevA.91.023615} {\bibfield  {journal} {\bibinfo  {journal} {Phys. Rev. A}\ }\textbf {\bibinfo {volume} {91}},\ \bibinfo {pages} {023615} (\bibinfo {year} {2015})}\BibitemShut {NoStop}%
\bibitem [{\citenamefont {Barenghi}\ \emph {et~al.}(2023)\citenamefont {Barenghi}, \citenamefont {Skrbek},\ and\ \citenamefont {Sreenivasan}}]{barenghi_skrbek_sreenivasan_2023}%
  \BibitemOpen
  \bibfield  {author} {\bibinfo {author} {\bibfnamefont {C.~F.}\ \bibnamefont {Barenghi}}, \bibinfo {author} {\bibfnamefont {L.}~\bibnamefont {Skrbek}},\ and\ \bibinfo {author} {\bibfnamefont {K.~R.}\ \bibnamefont {Sreenivasan}},\ }\href {https://doi.org/10.1017/9781009345651} {\emph {\bibinfo {title} {Quantum Turbulence}}}\ (\bibinfo  {publisher} {Cambridge University Press},\ \bibinfo {year} {2023})\BibitemShut {NoStop}%
\bibitem [{\citenamefont {Berloff}(2004)}]{Berloff2004Rarefaction}%
  \BibitemOpen
  \bibfield  {author} {\bibinfo {author} {\bibfnamefont {N.~G.}\ \bibnamefont {Berloff}},\ }\bibfield  {title} {\bibinfo {title} {Interactions of vortices with rarefaction solitary waves in a bose-einstein condensate and their role in the decay of superfluid turbulence},\ }\href {https://doi.org/10.1103/PhysRevA.69.053601} {\bibfield  {journal} {\bibinfo  {journal} {Phys. Rev. A}\ }\textbf {\bibinfo {volume} {69}},\ \bibinfo {pages} {053601} (\bibinfo {year} {2004})}\BibitemShut {NoStop}%
\bibitem [{\citenamefont {Kobayashi}\ and\ \citenamefont {Tsubota}(2007)}]{kobayashi2007quantum}%
  \BibitemOpen
  \bibfield  {author} {\bibinfo {author} {\bibfnamefont {M.}~\bibnamefont {Kobayashi}}\ and\ \bibinfo {author} {\bibfnamefont {M.}~\bibnamefont {Tsubota}},\ }\bibfield  {title} {\bibinfo {title} {Quantum turbulence in a trapped bose-einstein condensate},\ }\href {https://doi.org/10.1103/PhysRevA.76.045603} {\bibfield  {journal} {\bibinfo  {journal} {Phys. Rev. A}\ }\textbf {\bibinfo {volume} {76}},\ \bibinfo {pages} {045603} (\bibinfo {year} {2007})}\BibitemShut {NoStop}%
\bibitem [{\citenamefont {{\O}vereng}\ \emph {et~al.}(2025)\citenamefont {{\O}vereng}, \citenamefont {Baggaley},\ and\ \citenamefont {Angheluta}}]{overeng2025topological}%
  \BibitemOpen
  \bibfield  {author} {\bibinfo {author} {\bibfnamefont {V.}~\bibnamefont {{\O}vereng}}, \bibinfo {author} {\bibfnamefont {A.}~\bibnamefont {Baggaley}},\ and\ \bibinfo {author} {\bibfnamefont {L.}~\bibnamefont {Angheluta}},\ }\bibfield  {title} {\bibinfo {title} {Topological interactions in vortex-wave collisions in bose-einstein condensates},\ }\href@noop {} {\bibfield  {journal} {\bibinfo  {journal} {arXiv preprint arXiv:2510.01973}\ } (\bibinfo {year} {2025})}\BibitemShut {NoStop}%
\bibitem [{\citenamefont {Navon}\ \emph {et~al.}(2016)\citenamefont {Navon}, \citenamefont {Gaunt}, \citenamefont {Smith},\ and\ \citenamefont {Hadzibabic}}]{navon2016emergence}%
  \BibitemOpen
  \bibfield  {author} {\bibinfo {author} {\bibfnamefont {N.}~\bibnamefont {Navon}}, \bibinfo {author} {\bibfnamefont {A.~L.}\ \bibnamefont {Gaunt}}, \bibinfo {author} {\bibfnamefont {R.~P.}\ \bibnamefont {Smith}},\ and\ \bibinfo {author} {\bibfnamefont {Z.}~\bibnamefont {Hadzibabic}},\ }\bibfield  {title} {\bibinfo {title} {Emergence of a turbulent cascade in a quantum gas},\ }\href {https://doi.org/10.1038/nature20114} {\bibfield  {journal} {\bibinfo  {journal} {Nature}\ }\textbf {\bibinfo {volume} {539}},\ \bibinfo {pages} {72} (\bibinfo {year} {2016})}\BibitemShut {NoStop}%
\bibitem [{\citenamefont {Panico}\ \emph {et~al.}(2025)\citenamefont {Panico}, \citenamefont {Ciliberto}, \citenamefont {Martone}, \citenamefont {Congy}, \citenamefont {Ballarini}, \citenamefont {Lanotte},\ and\ \citenamefont {Pavloff}}]{Panico2025}%
  \BibitemOpen
  \bibfield  {author} {\bibinfo {author} {\bibfnamefont {R.}~\bibnamefont {Panico}}, \bibinfo {author} {\bibfnamefont {G.}~\bibnamefont {Ciliberto}}, \bibinfo {author} {\bibfnamefont {G.~I.}\ \bibnamefont {Martone}}, \bibinfo {author} {\bibfnamefont {T.}~\bibnamefont {Congy}}, \bibinfo {author} {\bibfnamefont {D.}~\bibnamefont {Ballarini}}, \bibinfo {author} {\bibfnamefont {A.~S.}\ \bibnamefont {Lanotte}},\ and\ \bibinfo {author} {\bibfnamefont {N.}~\bibnamefont {Pavloff}},\ }\bibfield  {title} {\bibinfo {title} {Topological pathways to two-dimensional quantum turbulence},\ }\href {https://doi.org/10.1103/PhysRevResearch.7.L022063} {\bibfield  {journal} {\bibinfo  {journal} {Phys. Rev. Res.}\ }\textbf {\bibinfo {volume} {7}},\ \bibinfo {pages} {L022063} (\bibinfo {year} {2025})}\BibitemShut {NoStop}%
\bibitem [{\citenamefont {Garcia-Orozco}\ \emph {et~al.}(2020)\citenamefont {Garcia-Orozco}, \citenamefont {Madeira}, \citenamefont {Galantucci}, \citenamefont {Barenghi},\ and\ \citenamefont {Bagnato}}]{Bagnato2020}%
  \BibitemOpen
  \bibfield  {author} {\bibinfo {author} {\bibfnamefont {A.~D.}\ \bibnamefont {Garcia-Orozco}}, \bibinfo {author} {\bibfnamefont {L.}~\bibnamefont {Madeira}}, \bibinfo {author} {\bibfnamefont {L.}~\bibnamefont {Galantucci}}, \bibinfo {author} {\bibfnamefont {C.~F.}\ \bibnamefont {Barenghi}},\ and\ \bibinfo {author} {\bibfnamefont {V.~S.}\ \bibnamefont {Bagnato}},\ }\bibfield  {title} {\bibinfo {title} {Intra-scales energy transfer during the evolution of turbulence in a trapped bose-einstein condensate},\ }\href {https://doi.org/10.1209/0295-5075/130/46001} {\bibfield  {journal} {\bibinfo  {journal} {Europhys. Lett.}\ }\textbf {\bibinfo {volume} {130}},\ \bibinfo {pages} {46001} (\bibinfo {year} {2020})}\BibitemShut {NoStop}%
\bibitem [{\citenamefont {Groszek}\ \emph {et~al.}(2020)\citenamefont {Groszek}, \citenamefont {Davis},\ and\ \citenamefont {Simula}}]{Groszek2020}%
  \BibitemOpen
  \bibfield  {author} {\bibinfo {author} {\bibfnamefont {A.~J.}\ \bibnamefont {Groszek}}, \bibinfo {author} {\bibfnamefont {M.~J.}\ \bibnamefont {Davis}},\ and\ \bibinfo {author} {\bibfnamefont {T.~P.}\ \bibnamefont {Simula}},\ }\bibfield  {title} {\bibinfo {title} {{Decaying quantum turbulence in a two-dimensional Bose-Einstein condensate at finite temperature}},\ }\href {https://doi.org/10.21468/SciPostPhys.8.3.039} {\bibfield  {journal} {\bibinfo  {journal} {SciPost Phys.}\ }\textbf {\bibinfo {volume} {8}},\ \bibinfo {pages} {039} (\bibinfo {year} {2020})}\BibitemShut {NoStop}%
\bibitem [{\citenamefont {Kanai}\ and\ \citenamefont {Guo}(2021)}]{Kanai2021}%
  \BibitemOpen
  \bibfield  {author} {\bibinfo {author} {\bibfnamefont {T.}~\bibnamefont {Kanai}}\ and\ \bibinfo {author} {\bibfnamefont {W.}~\bibnamefont {Guo}},\ }\bibfield  {title} {\bibinfo {title} {True mechanism of spontaneous order from turbulence in two-dimensional superfluid manifolds},\ }\href {https://doi.org/10.1103/PhysRevLett.127.095301} {\bibfield  {journal} {\bibinfo  {journal} {Phys. Rev. Lett.}\ }\textbf {\bibinfo {volume} {127}},\ \bibinfo {pages} {095301} (\bibinfo {year} {2021})}\BibitemShut {NoStop}%
\bibitem [{\citenamefont {Müller}\ and\ \citenamefont {Krstulovic}(2024)}]{muller2024}%
  \BibitemOpen
  \bibfield  {author} {\bibinfo {author} {\bibfnamefont {N.~P.}\ \bibnamefont {Müller}}\ and\ \bibinfo {author} {\bibfnamefont {G.}~\bibnamefont {Krstulovic}},\ }\bibfield  {title} {\bibinfo {title} {Exploring the equivalence between two-dimensional classical and quantum turbulence through velocity circulation statistics},\ }\href {https://doi.org/10.1103/PhysRevLett.132.094002} {\bibfield  {journal} {\bibinfo  {journal} {Phys. Rev. Lett.}\ }\textbf {\bibinfo {volume} {132}},\ \bibinfo {pages} {094002} (\bibinfo {year} {2024})}\BibitemShut {NoStop}%
\bibitem [{\citenamefont {Comaron}\ \emph {et~al.}(2025)\citenamefont {Comaron}, \citenamefont {Panico}, \citenamefont {Ballarini},\ and\ \citenamefont {Matuszewski}}]{comaron2025clustering}%
  \BibitemOpen
  \bibfield  {author} {\bibinfo {author} {\bibfnamefont {P.}~\bibnamefont {Comaron}}, \bibinfo {author} {\bibfnamefont {R.}~\bibnamefont {Panico}}, \bibinfo {author} {\bibfnamefont {D.}~\bibnamefont {Ballarini}},\ and\ \bibinfo {author} {\bibfnamefont {M.}~\bibnamefont {Matuszewski}},\ }\bibfield  {title} {\bibinfo {title} {Dynamics of onsager vortex clustering in decaying turbulent polariton quantum fluids},\ }\href {https://doi.org/10.1103/PhysRevResearch.7.L022006} {\bibfield  {journal} {\bibinfo  {journal} {Phys. Rev. Res.}\ }\textbf {\bibinfo {volume} {7}},\ \bibinfo {pages} {L022006} (\bibinfo {year} {2025})}\BibitemShut {NoStop}%
\bibitem [{\citenamefont {Ferrini}\ and\ \citenamefont {Koniakhin}(2025)}]{Ferrini2025}%
  \BibitemOpen
  \bibfield  {author} {\bibinfo {author} {\bibfnamefont {R.}~\bibnamefont {Ferrini}}\ and\ \bibinfo {author} {\bibfnamefont {S.~V.}\ \bibnamefont {Koniakhin}},\ }\bibfield  {title} {\bibinfo {title} {Driven-dissipative turbulence in exciton-polariton quantum fluids},\ }\href {https://doi.org/10.1103/khp5-5l3d} {\bibfield  {journal} {\bibinfo  {journal} {Phys. Rev. B}\ }\textbf {\bibinfo {volume} {112}},\ \bibinfo {pages} {205305} (\bibinfo {year} {2025})}\BibitemShut {NoStop}%
\bibitem [{\citenamefont {Estrecho}\ \emph {et~al.}(2019)\citenamefont {Estrecho}, \citenamefont {Gao}, \citenamefont {Bobrovska}, \citenamefont {Comber-Todd}, \citenamefont {Fraser}, \citenamefont {Steger}, \citenamefont {West}, \citenamefont {Pfeiffer}, \citenamefont {Levinsen}, \citenamefont {Parish}, \citenamefont {Liew}, \citenamefont {Matuszewski}, \citenamefont {Snoke}, \citenamefont {Truscott},\ and\ \citenamefont {Ostrovskaya}}]{Estrecho2019}%
  \BibitemOpen
  \bibfield  {author} {\bibinfo {author} {\bibfnamefont {E.}~\bibnamefont {Estrecho}}, \bibinfo {author} {\bibfnamefont {T.}~\bibnamefont {Gao}}, \bibinfo {author} {\bibfnamefont {N.}~\bibnamefont {Bobrovska}}, \bibinfo {author} {\bibfnamefont {D.}~\bibnamefont {Comber-Todd}}, \bibinfo {author} {\bibfnamefont {M.~D.}\ \bibnamefont {Fraser}}, \bibinfo {author} {\bibfnamefont {M.}~\bibnamefont {Steger}}, \bibinfo {author} {\bibfnamefont {K.}~\bibnamefont {West}}, \bibinfo {author} {\bibfnamefont {L.~N.}\ \bibnamefont {Pfeiffer}}, \bibinfo {author} {\bibfnamefont {J.}~\bibnamefont {Levinsen}}, \bibinfo {author} {\bibfnamefont {M.~M.}\ \bibnamefont {Parish}}, \bibinfo {author} {\bibfnamefont {T.~C.~H.}\ \bibnamefont {Liew}}, \bibinfo {author} {\bibfnamefont {M.}~\bibnamefont {Matuszewski}}, \bibinfo {author} {\bibfnamefont {D.~W.}\ \bibnamefont {Snoke}}, \bibinfo {author} {\bibfnamefont {A.~G.}\ \bibnamefont {Truscott}},\ and\ \bibinfo {author} {\bibfnamefont {E.~A.}\ \bibnamefont {Ostrovskaya}},\ }\bibfield
  {title} {\bibinfo {title} {Direct measurement of polariton-polariton interaction strength in the thomas-fermi regime of exciton-polariton condensation},\ }\href {https://doi.org/10.1103/PhysRevB.100.035306} {\bibfield  {journal} {\bibinfo  {journal} {Phys. Rev. B}\ }\textbf {\bibinfo {volume} {100}},\ \bibinfo {pages} {035306} (\bibinfo {year} {2019})}\BibitemShut {NoStop}%
\bibitem [{\citenamefont {Panico}\ \emph {et~al.}(2023{\natexlab{b}})\citenamefont {Panico}, \citenamefont {Lanotte}, \citenamefont {Trypogeorgos}, \citenamefont {Gigli}, \citenamefont {De~Giorgi}, \citenamefont {Sanvitto},\ and\ \citenamefont {Ballarini}}]{conformal2023}%
  \BibitemOpen
  \bibfield  {author} {\bibinfo {author} {\bibfnamefont {R.}~\bibnamefont {Panico}}, \bibinfo {author} {\bibfnamefont {A.~S.}\ \bibnamefont {Lanotte}}, \bibinfo {author} {\bibfnamefont {D.}~\bibnamefont {Trypogeorgos}}, \bibinfo {author} {\bibfnamefont {G.}~\bibnamefont {Gigli}}, \bibinfo {author} {\bibfnamefont {M.}~\bibnamefont {De~Giorgi}}, \bibinfo {author} {\bibfnamefont {D.}~\bibnamefont {Sanvitto}},\ and\ \bibinfo {author} {\bibfnamefont {D.}~\bibnamefont {Ballarini}},\ }\bibfield  {title} {\bibinfo {title} {{Conformal invariance of 2D quantum turbulence in an exciton–polariton fluid of light}},\ }\href {https://doi.org/10.1063/5.0167655} {\bibfield  {journal} {\bibinfo  {journal} {Applied Physics Reviews}\ }\textbf {\bibinfo {volume} {10}},\ \bibinfo {pages} {041418} (\bibinfo {year} {2023}{\natexlab{b}})}\BibitemShut {NoStop}%
\bibitem [{Pro(2017)}]{Proukakis_Snoke_Littlewood_2017}%
  \BibitemOpen
  \href@noop {} {\emph {\bibinfo {title} {Universal Themes of Bose-Einstein Condensation}}}\ (\bibinfo  {publisher} {Cambridge University Press},\ \bibinfo {year} {2017})\BibitemShut {NoStop}%
\bibitem [{\citenamefont {Barenghi}\ and\ \citenamefont {Parker}(2016)}]{Barenghi_book}%
  \BibitemOpen
  \bibfield  {author} {\bibinfo {author} {\bibfnamefont {C.~F.}\ \bibnamefont {Barenghi}}\ and\ \bibinfo {author} {\bibfnamefont {N.~G.}\ \bibnamefont {Parker}},\ }\href {https://link.springer.com/book/10.1007/978-3-319-42476-7} {\emph {\bibinfo {title} {A primer on Quantum Fluids}}}\ (\bibinfo  {publisher} {Springer},\ \bibinfo {year} {2016})\BibitemShut {NoStop}%
\bibitem [{SI()}]{SI}%
  \BibitemOpen
  \href@noop {} {\bibinfo {title} {{\rm See Supplementary Information, which contains the following references: \cite{Comaron2021,weiler2008spontaneous,Rooney2013,Comaron2019} }}}\BibitemShut {NoStop}%
\bibitem [{Note1()}]{Note1}%
  \BibitemOpen
  \bibinfo {note} {Note that $C_{\max }$ is always attained with at least approximately ten vortices still remaining, and that the enhancement with respect to the conservative case largely exceeds the statistical uncertainty of the ensemble average.}\BibitemShut {Stop}%
\bibitem [{Note2()}]{Note2}%
  \BibitemOpen
  \bibinfo {note} {Although vortex decay exhibits different dynamical regimes, we extract $\tau _{n_\protect \mathrm {v}}$ by fitting the vortex-number evolution with a stretched exponential $N_v(t)\sim \exp [-(t/\tau _{n_\protect \mathrm {v}})^\alpha ]$, treating both $\tau _{n_\protect \mathrm {v}}$ and $\alpha $ as free parameters.}\BibitemShut {Stop}%
\bibitem [{\citenamefont {Gladilin}\ and\ \citenamefont {Wouters}(2017)}]{Gladilin_2017}%
  \BibitemOpen
  \bibfield  {author} {\bibinfo {author} {\bibfnamefont {V.~N.}\ \bibnamefont {Gladilin}}\ and\ \bibinfo {author} {\bibfnamefont {M.}~\bibnamefont {Wouters}},\ }\bibfield  {title} {\bibinfo {title} {Interaction and motion of vortices in nonequilibrium quantum fluids},\ }\href {https://doi.org/10.1088/1367-2630/aa83a1} {\bibfield  {journal} {\bibinfo  {journal} {New Journal of Physics}\ }\textbf {\bibinfo {volume} {19}},\ \bibinfo {pages} {105005} (\bibinfo {year} {2017})}\BibitemShut {NoStop}%
\bibitem [{\citenamefont {Gladilin}\ and\ \citenamefont {Wouters}(2019)}]{Gladilin_2019}%
  \BibitemOpen
  \bibfield  {author} {\bibinfo {author} {\bibfnamefont {V.~N.}\ \bibnamefont {Gladilin}}\ and\ \bibinfo {author} {\bibfnamefont {M.}~\bibnamefont {Wouters}},\ }\bibfield  {title} {\bibinfo {title} {Multivortex states and dynamics in nonequilibrium polariton condensates},\ }\href {https://doi.org/10.1088/1751-8121/ab3abc} {\bibfield  {journal} {\bibinfo  {journal} {Journal of Physics A: Mathematical and Theoretical}\ }\textbf {\bibinfo {volume} {52}},\ \bibinfo {pages} {395303} (\bibinfo {year} {2019})}\BibitemShut {NoStop}%
\bibitem [{\citenamefont {Wachtel}\ \emph {et~al.}(2016)\citenamefont {Wachtel}, \citenamefont {Sieberer}, \citenamefont {Diehl},\ and\ \citenamefont {Altman}}]{wachtel2016electrodynamic}%
  \BibitemOpen
  \bibfield  {author} {\bibinfo {author} {\bibfnamefont {G.}~\bibnamefont {Wachtel}}, \bibinfo {author} {\bibfnamefont {L.~M.}\ \bibnamefont {Sieberer}}, \bibinfo {author} {\bibfnamefont {S.}~\bibnamefont {Diehl}},\ and\ \bibinfo {author} {\bibfnamefont {E.}~\bibnamefont {Altman}},\ }\bibfield  {title} {\bibinfo {title} {{Electrodynamic duality and vortex unbinding in driven-dissipative condensates}},\ }\href {https://doi.org/10.1103/PhysRevB.94.104520} {\bibfield  {journal} {\bibinfo  {journal} {Phys. Rev. B}\ }\textbf {\bibinfo {volume} {94}},\ \bibinfo {pages} {104520} (\bibinfo {year} {2016})}\BibitemShut {NoStop}%
\bibitem [{\citenamefont {Comaron}\ \emph {et~al.}(2021)\citenamefont {Comaron}, \citenamefont {Carusotto}, \citenamefont {Szyma{\'{n}}ska},\ and\ \citenamefont {Proukakis}}]{Comaron2021}%
  \BibitemOpen
  \bibfield  {author} {\bibinfo {author} {\bibfnamefont {P.}~\bibnamefont {Comaron}}, \bibinfo {author} {\bibfnamefont {I.}~\bibnamefont {Carusotto}}, \bibinfo {author} {\bibfnamefont {M.~H.}\ \bibnamefont {Szyma{\'{n}}ska}},\ and\ \bibinfo {author} {\bibfnamefont {N.~P.}\ \bibnamefont {Proukakis}},\ }\bibfield  {title} {\bibinfo {title} {Non-equilibrium berezinskii-kosterlitz-thouless transition in driven-dissipative condensates (a)},\ }\href {https://doi.org/10.1209/0295-5075/133/17002} {\bibfield  {journal} {\bibinfo  {journal} {{EPL} (Europhysics Letters)}\ }\textbf {\bibinfo {volume} {133}},\ \bibinfo {pages} {17002} (\bibinfo {year} {2021})}\BibitemShut {NoStop}%
\bibitem [{\citenamefont {Weiler}\ \emph {et~al.}(2008)\citenamefont {Weiler}, \citenamefont {Neely}, \citenamefont {Scherer}, \citenamefont {Bradley}, \citenamefont {Davis},\ and\ \citenamefont {Anderson}}]{weiler2008spontaneous}%
  \BibitemOpen
  \bibfield  {author} {\bibinfo {author} {\bibfnamefont {C.~N.}\ \bibnamefont {Weiler}}, \bibinfo {author} {\bibfnamefont {T.~W.}\ \bibnamefont {Neely}}, \bibinfo {author} {\bibfnamefont {D.~R.}\ \bibnamefont {Scherer}}, \bibinfo {author} {\bibfnamefont {A.~S.}\ \bibnamefont {Bradley}}, \bibinfo {author} {\bibfnamefont {M.~J.}\ \bibnamefont {Davis}},\ and\ \bibinfo {author} {\bibfnamefont {B.~P.}\ \bibnamefont {Anderson}},\ }\bibfield  {title} {\bibinfo {title} {Spontaneous vortices in the formation of bose--einstein condensates},\ }\href {https://doi.org/10.1038/nature07334} {\bibfield  {journal} {\bibinfo  {journal} {Nature}\ }\textbf {\bibinfo {volume} {455}},\ \bibinfo {pages} {948} (\bibinfo {year} {2008})}\BibitemShut {NoStop}%
\bibitem [{\citenamefont {Rooney}\ \emph {et~al.}(2013)\citenamefont {Rooney}, \citenamefont {Neely}, \citenamefont {Anderson},\ and\ \citenamefont {Bradley}}]{Rooney2013}%
  \BibitemOpen
  \bibfield  {author} {\bibinfo {author} {\bibfnamefont {S.~J.}\ \bibnamefont {Rooney}}, \bibinfo {author} {\bibfnamefont {T.~W.}\ \bibnamefont {Neely}}, \bibinfo {author} {\bibfnamefont {B.~P.}\ \bibnamefont {Anderson}},\ and\ \bibinfo {author} {\bibfnamefont {A.~S.}\ \bibnamefont {Bradley}},\ }\bibfield  {title} {\bibinfo {title} {Persistent-current formation in a high-temperature bose-einstein condensate: An experimental test for classical-field theory},\ }\href {https://doi.org/10.1103/PhysRevA.88.063620} {\bibfield  {journal} {\bibinfo  {journal} {Phys. Rev. A}\ }\textbf {\bibinfo {volume} {88}},\ \bibinfo {pages} {063620} (\bibinfo {year} {2013})}\BibitemShut {NoStop}%
\bibitem [{\citenamefont {Comaron}\ \emph {et~al.}(2019)\citenamefont {Comaron}, \citenamefont {Larcher}, \citenamefont {Dalfovo},\ and\ \citenamefont {Proukakis}}]{Comaron2019}%
  \BibitemOpen
  \bibfield  {author} {\bibinfo {author} {\bibfnamefont {P.}~\bibnamefont {Comaron}}, \bibinfo {author} {\bibfnamefont {F.}~\bibnamefont {Larcher}}, \bibinfo {author} {\bibfnamefont {F.}~\bibnamefont {Dalfovo}},\ and\ \bibinfo {author} {\bibfnamefont {N.~P.}\ \bibnamefont {Proukakis}},\ }\bibfield  {title} {\bibinfo {title} {Quench dynamics of an ultracold two-dimensional bose gas},\ }\href {https://doi.org/10.1103/PhysRevA.100.033618} {\bibfield  {journal} {\bibinfo  {journal} {Phys. Rev. A}\ }\textbf {\bibinfo {volume} {100}},\ \bibinfo {pages} {033618} (\bibinfo {year} {2019})}\BibitemShut {NoStop}%
\end{thebibliography}
\end{document}